\documentclass[aps,pra,reprint,floatfix,superscriptaddress]{revtex4-2}
\usepackage{numprint}
\npdecimalsign{.}
\usepackage{amsmath}
\usepackage{graphicx}
\usepackage{tabularx}
\graphicspath{{plots/}}
\usepackage{isotope}
\usepackage{physics}
\usepackage{nicefrac}
\usepackage{braket}
\usepackage{siunitx}
\begin{document}

\title[Article Title]{Precision continuous-wave laser measurement of the $\text{1}^\text{3}\text{S}_\text{1} \to \text{2}^\text{3}\text{S}_\text{1}$ interval in positronium }

\author{Lucas de Sousa Borges}
\email{lucas.borges@phys.ethz.ch}
\affiliation{Institute for Particle Physics and Astrophysics, ETH Z\"urich, CH-8093 Z\"urich, Switzerland}

\author{Edward Thorpe-Woods}
\email{ethorpe@ethz.ch}
\affiliation{Institute for Particle Physics and Astrophysics, ETH Z\"urich, CH-8093 Z\"urich, Switzerland}

\author{Evans Javary}
\email{ejavary@ethz.ch}
\affiliation{Institute for Particle Physics and Astrophysics, ETH Z\"urich, CH-8093 Z\"urich, Switzerland}

\author{Paolo Crivelli}
\email{crivelli@phys.ethz.ch}
\affiliation{Institute for Particle Physics and Astrophysics, ETH Z\"urich, CH-8093 Z\"urich, Switzerland}

\date{\today}

\begin{abstract}
We report a 4.9\,ppb measurement of the positronium $\text{1}^\text{3}\text{S}_\text{1} \to \text{2}^\text{3}\text{S}_\text{1}$ interval using continuous-wave two-photon laser spectroscopy. The transition is detected via photoionization by the same excitation laser. The resulting positrons are guided to a microchannel plate detector, surrounded by scintillators to detect the annihilation photons in coincidence, thereby reducing the background. A Monte Carlo lineshape simulation, accounting for effects such as the second-order Doppler shift and the AC Stark shift, is used to extract a transition frequency of $\numprint{1233607224.1}(6.0)\,\text{MHz}$, consistent with the previous 2.6\,ppb determination of this transition and with the most recent QED calculations at order $\mathcal{O}(\alpha^7\ln^2(1/\alpha))$, which predict $\numprint{1233607222.12}(58)\,\text{MHz}$. Combining the two measurements gives $\numprint{1233607218.1}(2.8)\,\text{MHz}$, reducing the tension with QED to about $1.4\,\sigma$. We also present a semi-analytical lineshape model of $\text{1}^\text{3}\text{S}_\text{1} \to \text{2}^\text{3}\text{S}_\text{1}$ of positronium, which shows excellent agreement with detailed simulations and is validated by the experimental data. This expands on previous work with stable atoms by incorporating effects such as limited lifetime of the atoms, photoionization and AC Stark shift. The lineshape modelling is also applicable to other unstable systems, such as muonium. This provides a powerful tool for optimizing the experimental parameters and gaining deeper insights without the need for computationally intensive simulations.

\end{abstract}

\maketitle

\section{\label{sec:introduction}Introduction}

Positronium (Ps), the bound state of an electron and a positron, provides a unique laboratory for testing bound-state QED and search for new physics (see \cite{Adkins:2022omi} and references therein for a recent comprehensive review on the field). Owing to the absence of nuclear structure, it allows direct comparison between precision measurements of energy intervals and high-order theoretical predictions reaching sub-ppb accuracy \cite{Czarnecki1999, Baker2014, Adkins2014, Eides2014, Adkins2014a, Eides2015, Adkins2015, Adkins2015a, Adkins2016, Eides2016, Eides2017, Adkins:2022coe, Eides2021, Eides2022, Eides2023}. 

Over the past decades, precision measurements of several Ps properties have been performed. The long-standing orthopositronium lifetime puzzle was resolved, yielding excellent agreement with QED predictions~\cite{Vallery:2003iz,Jinnouchi:2003hr}. However, the experimental precision still lags behind the theoretical determination~\cite{Adkins:2022omi} by about two orders of magnitude, calling for further improvement. 

The ground-state hyperfine splitting has been measured with reduced systematics at Tokyo~\cite{Ishida:2013waa}, while the $n=2$ fine-structure intervals have been determined with high precision through microwave spectroscopy~\cite{Gurung:2020hms}. 

The first demonstration of laser cooling of positronium holds great promise for producing colder samples suitable for next-generation high-precision spectroscopy~\cite{PhysRevLett.132.083402,Shu2024}. 

Positronium also serves as a sensitive probe of fundamental symmetries, enabling searches for CP and CPT violation~\cite{Yamazaki:2009hp,Moskal:2024jfu} and for new physics beyond the Standard Model, such as dark-sector particles or new light bosons~\cite{Frugiuele_2019,Vigo:2019bou}. 

Renewed efforts toward measuring the $\text{1}^\text{3}\text{S}_\text{1} \to \text{2}^\text{3}\text{S}_\text{1}$ interval are ongoing~\cite{Heiss2025}, building upon the seminal work of Fee \emph{et al.}~\cite{Fee1993}. In this work, we report a new precision measurement at a level comparable to the current best experimental determination of this interval.

\section{\label{sec:experimental}Experimental methods}

\begin{figure*}
\includegraphics[width=0.7\textwidth]{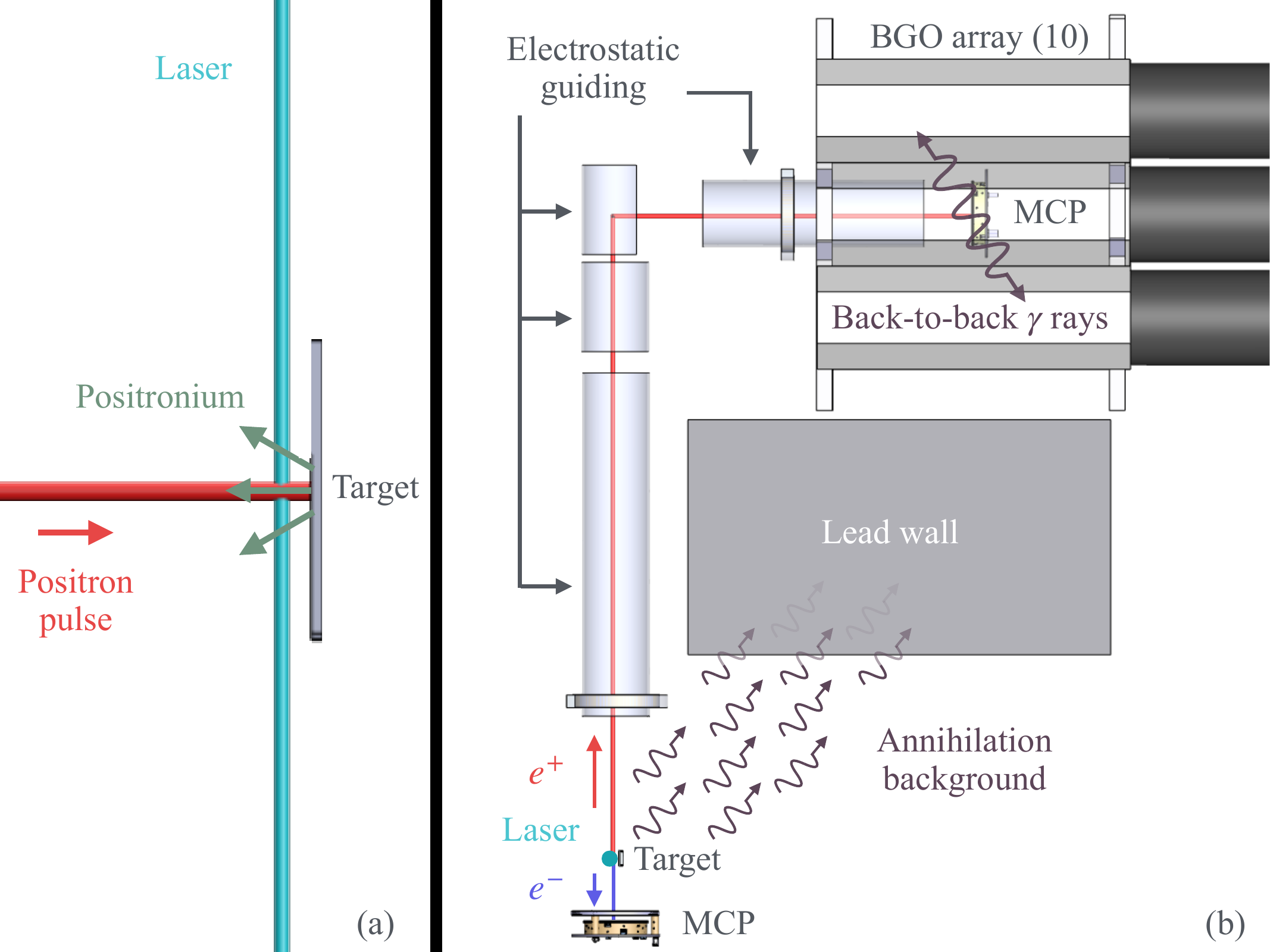}
\caption[Sketch of the experimental scheme for positronium $\text{1}^\text{3}\text{S}_\text{1} \to \text{2}^\text{3}\text{S}_\text{1}$ spectroscopy.]{Sketch of the experimental scheme for positronium $\text{1}^\text{3}\text{S}_\text{1} \to \text{2}^\text{3}\text{S}_\text{1}$ spectroscopy. (a) A positron pulse impacts a porous target and forms positronium. The atoms are then emitted in vacuum and cross a CW laser at \SI{486}{\nano\meter}. The laser excites the atoms to the 2S state, which are ionised by another photon absorption. (b) The photoionized components of the atom ($e^+$, $e^-$) are separated by carefully selected voltages on the target, bottom MCP grid and electrostatic lenses.
The photoionized positron is guided toward the detection region, spatially separated from the formation region to avoid annihilation background from the initial positron pulse. Lead shielding between the formation and detection regions further suppresses this background. The guided positron finally hits a microchannel-plate detector and annihilates, producing two back-to-back \SI{511}{\kilo\electronvolt} $\gamma$-rays that are detected in an array of ten calibrated BGO scintillators.}
\label{fig:detection-scheme}
\end{figure*}

\begin{figure}
    \centering
\includegraphics[width=0.75\linewidth]{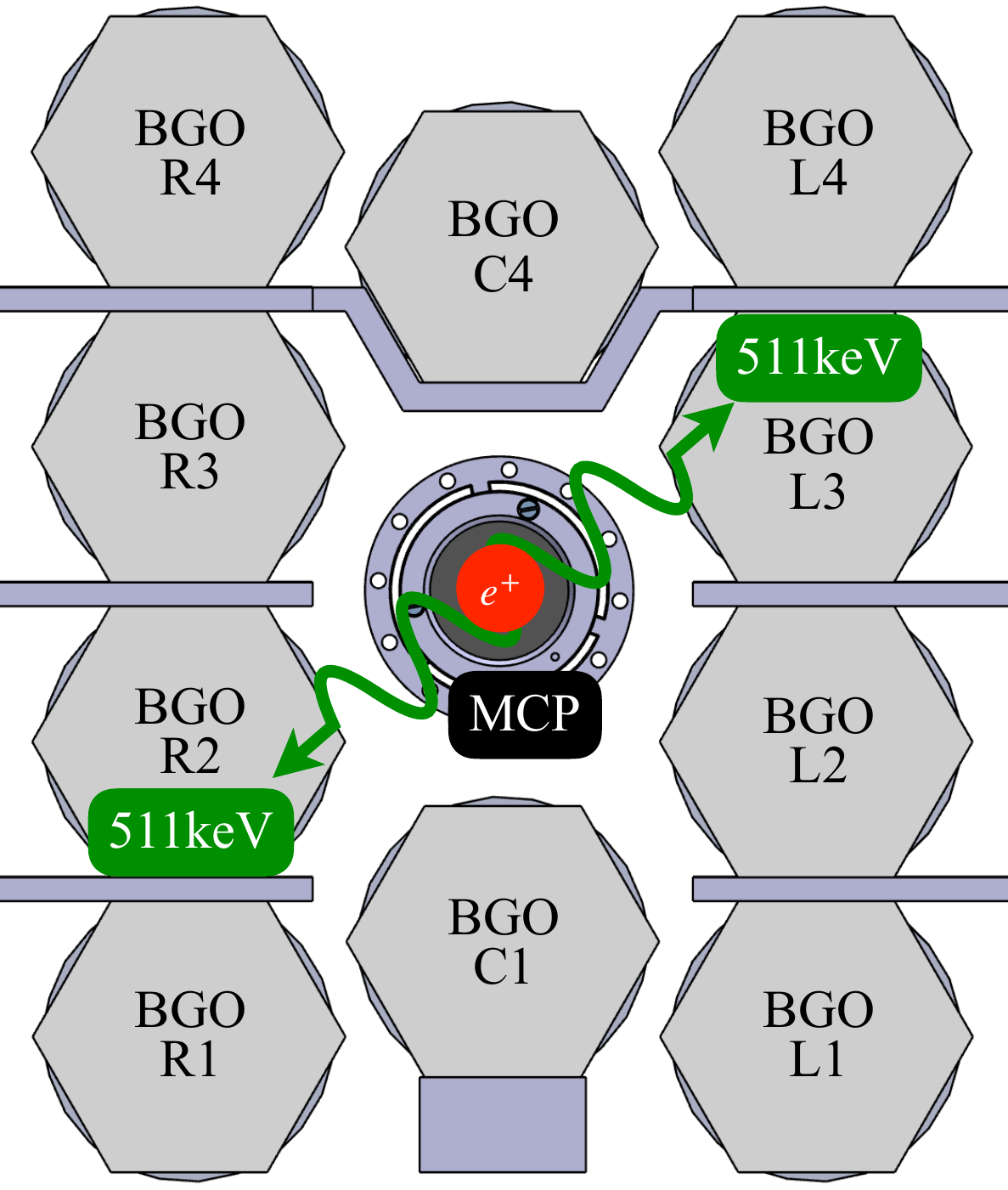}
    \caption{MCP surrounded by BGO scintillators for the triple coincidence detection method (see text for more details).}
    \label{fig:MCP+BGOs}
\end{figure}

Positrons ($e^+$) from a neon-moderated $^{22}$Na radioactive source \cite{doi:10.1063/1.97441} are accumulated in a buffer gas trap \cite{PhysRevA.46.5696} and pulsed towards a porous silica converter to generate positronium atoms \cite{Crivelli2010,Cassidy2010} after being extracted in a magnetic field-free region \cite{Cooke_2016}. 
The $\text{1}^\text{3}\text{S}_\text{1} \to \text{2}^\text{3}\text{S}_\text{1}$ transition in positronium is dipole-forbidden, but it can be excited through the absorption of two counter-propagating photons at \SI{486}{\nano\meter}, which cancels the first-order Doppler shift. We drive the excitation using continuous-wave (CW) laser light enhanced in a Fabry–Perot optical cavity, which ensures a high circulating power and enables efficient two-photon excitation despite the small cross section of the process and the limited interaction time between the atoms and the laser.  
Detection is achieved via photoionization of the excited atoms by absorption of an additional \SI{486}{\nano\meter} photon from the same laser. The photoionized positrons are then electrostatically guided to a low-background detection region, which is shielded by a lead wall from the $\gamma$-rays produced by positron annihilations in the target (see Fig.~\ref{fig:detection-scheme}). This shielding allows the detection of single photoionized positrons on a microchannel plate (MCP) in coincidence with the reconstruction of the two back-to-back \SI{511}{\kilo\electronvolt} $\gamma$-rays from the positron annihilation, recorded by an array of ten BGO scintillators placed around the MCP (see Fig.~\ref{fig:MCP+BGOs}).

\subsection{\label{subsec:positronium_formation} Positronium formation}
The pulsed positron beam used for positronium production has been described in detail in \cite{Cooke_2016,PhysRevA.111.012810,Heiss:2021qxv}. With a positron source activity of \SI{53(1)}{\mega\becquerel}, about $8000$ positrons per second at \SI{4.7}{\kilo\electronvolt} in \SI{5}{\nano\second} bunches impinge on a porous silica target to form positronium in the triplet spin state (ortho-positronium) with a conversion efficiency of 20(5)\% \cite{Crivelli2010, Cassidy2010}. From previous studies, this Ps production method has been shown to emit the atoms in vacuum with a velocity distribution that can be parametrised by a Maxwell-Boltzmann distribution at 600\si{\kelvin} for the given positron implantation energy \cite{Deller2015, PhysRevA.111.012810}.
    
\subsection{\label{subsec:laser} Laser system}

\begin{figure*}
\includegraphics[width=0.9\textwidth]{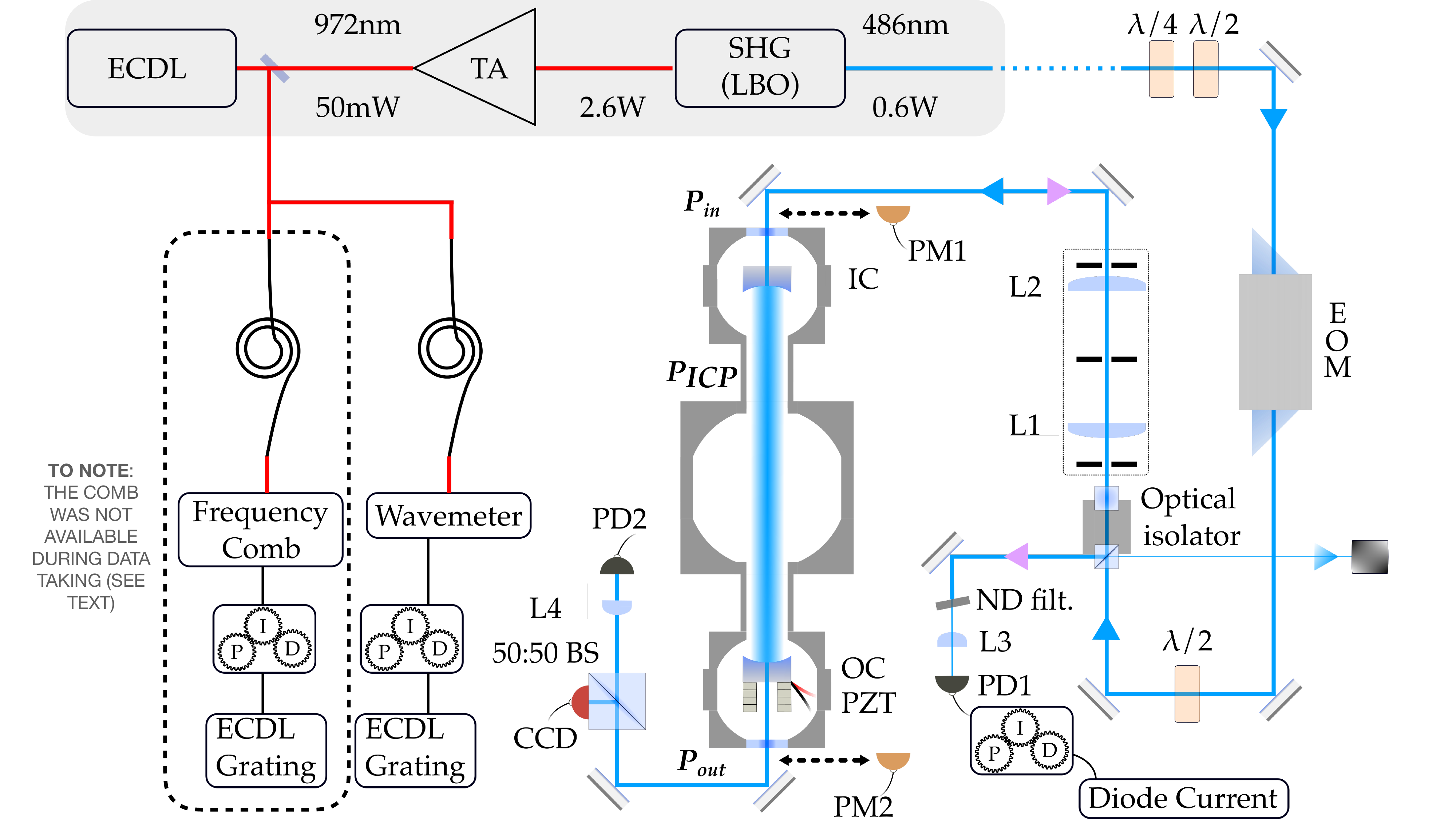}
    \caption[Sketch of the laser system.]{Sketch of the laser system. Extended cavity diode laser (ECDL), tapered amplifier (TA), second harmonic generation (SHG), lithium triborate (LBO), beam splitter (BS), electro-optic modulator (EOM), mode matching lenses (L1-L2), photodiode (PD), power monitor (PM), input coupler (IC), output coupler (OC),  piezoelectric transducer stack (PZT), neutral density filter (ND filter.), ultra-low expansion cavity (ULE).} 
    \label{fig:optical_line_CW}
\end{figure*}

Figure \ref{fig:optical_line_CW} shows an overview of the laser system used for the measurement. The grey shaded region corresponds to a commercial Toptica DL pro with a frequency-doubled output at \SI{486}{\nano\meter}, characterised by a \SI{100}{\kilo\hertz} linewidth. The output of the laser is amplified in an enhancement cavity consisting of two identical concave ($R=\SI{2}{\meter}$) dielectric mirrors with a transmission of 7.0(1) ppm, locked via the Pound-Drever-Hall (PDH) method \cite{Black2001}. The EOM generates \SI{17.8}{\mega\hertz} sidebands required for the locking scheme; the laser is mode-matched to the cavity by lenses L1 and L2; the input power in the cavity is controlled via a $\lambda$/2 plate in front of the optical isolator. The reflected power is guided towards the PDH photodiode to generate the error signal. The coupling efficiency reaches 30\%, limited by impedance-mismatch of the cavity mirrors, with a buildup factor in the range of \numprint{20000}--\numprint{25000}, meaning an input power of \SIrange{60}{75}{\milli\watt} is sufficient to keep a circulating power of \SIrange{400}{500}{\watt}. Higher powers up to \SI{3.5}{\kilo\watt} are attainable, but the dielectric mirrors degrade within minutes, whereas the cavity performance lasts several weeks at \SI{500}{\watt}. The frequency of the laser is kept stable at a level of \SI{1}{\mega\hertz} at \SI{486}{\nano\meter} with a wavemeter (High Finesse WS7-60). The frequency comb was being upgraded and was therefore unavailable during this data-taking period. The PID output of the device acts on the piezoelectric element voltage of the laser diode grating to stabilise the frequency. The frequency metrology procedure is detailed in appendix \ref{sec:appendix_frequency_metrology} and the total laser frequency uncertainty amounts to \SI{5.4}{\mega\hertz}. The cavity focal point is located in front of the target with a waist size $w_0=\SI{310(2)}{\micro\meter}$ located \SI{3.5(0.5)}{\milli\meter} from the target.  The photoionization probability is estimated with Monte Carlo simulations to be $6(2)\cdot10^{-6}$ on resonance.\\

\subsection{\label{subsec:detection} Signal detection}

The guiding system consists of a set of electrodes and voltages applied to the target and MCPs. The electric field seen by the atom upon excitation is of \SI{340(10)}{\volt\per\meter}. Simulations performed with SIMION~\cite{DAHL20003} indicate a 95(1)\% guiding efficiency. The efficiency for the BGO scintillators to detect the back-to-back \SI{511}{\kilo\electronvolt} annihilation photons was estimated with simulations to be of $65$\%. The different efficiencies of the processes described are summarised in Table \ref{tab:detection_efficiency}. The uncertainty on physical parameters such as velocity distribution, laser power, and laser position results in an uncertainty on the photoionization probability. The expected rate considering positron pulses of 4100(900) $e^+$ is of $0.0015(8)$ events per pulse.

\begin{table}[h]
\centering
\begin{tabular}{lr}
\hline\hline
\textbf{} & \textbf{Efficiency}  \\
\hline
Positronium formation    &   0.20(5)   \\
Excitation \& photoionization & $6(2)\cdot10^{-6}$  \\
Photoionized positron guiding    & 0.95(1)   \\
Positron detection in MCP  &  0.5(1)   \\
Annihilation $\gamma$ reconstruction with BGOs     &   0.65(5) \\
\hline
\rule{0pt}{2.5ex} Total           & $4(2)\cdot10^{-7}$ \\
\hline\hline
\end{tabular}
\caption{Summary of different efficiencies, giving the total probability of detecting a signal event per incoming positron.}
\label{tab:detection_efficiency}
\end{table}

\subsection{\label{subsec:DAQ_analysis} DAQ and analysis}
The data are acquired with a WaveDREAM2 Rev. E~\cite{GALLI2019399}, recording a \SI{1}{\micro\second} window at 1 GS/s (\SI{1}{\nano\second} resolution). The waveform of the MCP, each BGO, the lead tungstate (PbWO$_4$) crystal (see next paragraph) and the cavity photodiode are recorded and analysed in post-processing.\\

The laser frequency is changed every 20 minutes. The analysis is performed on an event-by-event basis. The positron moderator degradation and regrowth every 8 hours cause a variation in the amount of positronium atoms in the experimental chamber. This effect is corrected by recording the total amount of positrons impinging on target with a PbWO$_4$ scintillator. A photodiode monitors the output of the cavity, allowing for correct counting of the number of events where the laser cavity was locked and had a laser power within a range of \SIrange{400}{500}{\watt}.\\

To suppress the background, a triple coincidence detection scheme is used. With the initial timing $t_0$ given by the detection of the positron pulse in the lead tungstate scintillator, a photoionized positron is expected to arrive at the MCP within a time-of-flight window of \SIrange{85}{135}{\nano\second}. In this time frame, if a hit is recorded in the MCP in coincidence with the detection of two back-to-back \SI{511}{\kilo\electronvolt} $\gamma$-rays, as depicted in Fig. \ref{fig:MCP+BGOs}, the event is counted as a signal event. \\

\section{Lineshape modelling} 

\subsection{Monte Carlo simulation for lineshape fitting}\label{sec:MCsimulation} 
\begin{table}[t]
\centering
\setlength{\tabcolsep}{10pt}
\begin{tabular}{lcl}
\hline\hline
\textbf{Parameter} & \textbf{Unit} & \textbf{Value} \\
\hline
\textbf{Beam creation on target} & mm & 2.0(2) \\
\textbf{Distance laser-target} & mm & 3.5(5) \\
\textbf{Laser beam waist }& $\mu$m & 310(2) \\
\textbf{Laser power} & W & 460(50) \\
\textbf{Mean temperature }& K & 600(30) \\
\hline\hline
\end{tabular}
\caption{Experimental parameters and their uncertainty used as an input to the simulation.}
\label{tab:exp_parameters}
\end{table}

To fit the data the lineshape is obtained from a detailed Monte Carlo model of the $\text{1}^\text{3}\text{S}_\text{1} \to \text{2}^\text{3}\text{S}_\text{1}$ transition validated with previous measurements \cite{PhysRevA.111.012810,Borges:2025}. The simulation reproduces the experimental conditions, including the velocity and spatial distributions of the positronium atoms and the laser parameters, and incorporates effects such as the second-order Doppler and AC Stark shifts.
The initial Ps spatial distribution is modelled as a two-dimensional Gaussian with measured standard deviation \SI{2.0(2)}{\milli\meter}, corresponding to the positron beam size on the target. The mean implantation depth of 4.7 keV SiO$_2$ is only a few hundred nm which is negligible compared to the beam diameter. We therefore take the initial Ps positions to follow the beam profile rather than being broadened by diffusion in the porous layer. The velocity of the atoms can be parametrised with a Maxwell–Boltzmann distribution characterized by a typical temperature of 600(30) K at the given positron implantation energy, and their emission directions are drawn according to a $cos(\theta)$  distribution, where $\theta$ is the angle with respect to the normal to the target surface \cite{PhysRevA.111.012810, Heiss:2021qxv}.
Additional inputs to the simulation are the laser--target distance of \SI{3.5(5)}{\milli\meter} and the laser power of \SI{460(50)}{\watt}.

For each such parameter, the simulation is repeated with the input varied within its quoted experimental uncertainty, and the corresponding change in the extracted line center is taken as a systematic uncertainty (see Sec. \ref{sec:systematics}). 
 
\begin{figure}[h!]
\includegraphics[width=\columnwidth]{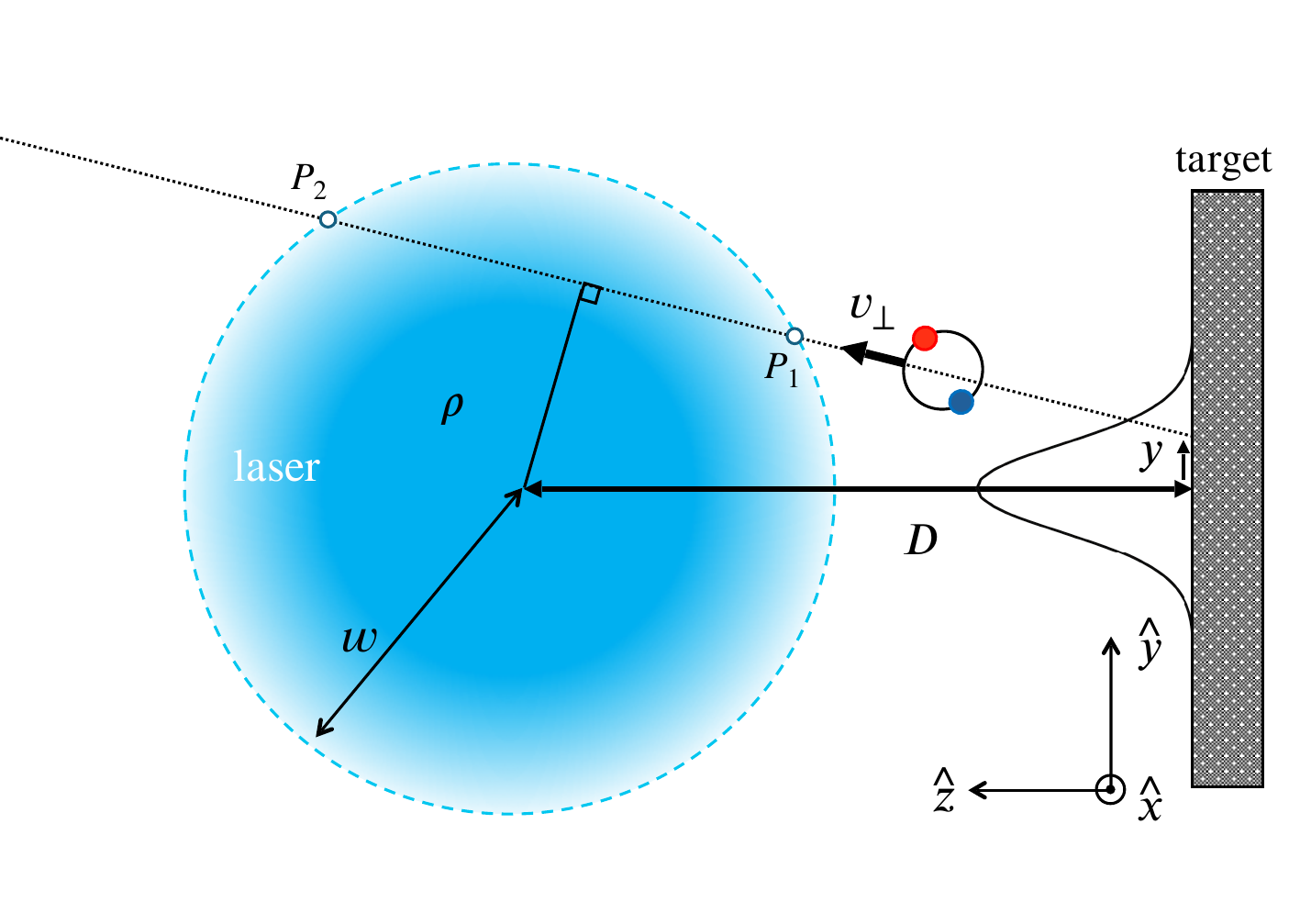} 
   \caption[Line] {Ps trajectory as it traverses the laser beam. The light field grows as Ps moves towards the beam, passing at its distance of closest approach of the beam axis, $\rho$.
\label{fig:laser_transit}}
\end{figure}

Each atom is initially generated in the 1S ground state. The interaction between the atoms and the continuous-wave (CW) laser field at 486 nm is described using the optical Bloch equations integrated numerically with a fourth-order Runge–Kutta method (see Appendix B of \cite{PhysRevA.111.012810} for the details). The intersections of the atoms with the laser field are calculated using ray-cylinder intersection method from atoms trajectories, laser position and waist (see Fig. \ref{fig:laser_transit}). This enables the temporal evolution of the atomic state to be precisely followed throughout its interaction with the laser field, including the excitation to the 2S state and the possible photoionization upon absorption of an additional photon from the same laser field.

\subsection{Semi-analytical lineshape} \label{sec: MC Simulation}
The Monte Carlo simulations described above fully model the experimental conditions. However, these simulations are computationally intensive, particularly when a large number of simulated spectra is required to fit the experimental data. To reduce dependence on simulations and provide a clearer mathematical description of the $\text{1}^\text{3}\text{S}_\text{1} \to \text{2}^\text{3}\text{S}_\text{1}$ positronium spectrum, we have developed a semi-analytical lineshape model based on second-order perturbation theory.

The mathematical details of the lineshape are derived in detail in appendix \ref{sec:appendix-lineshape}. The function for the 2S and ionisation lineshape are given in equations \ref{eq: explicit2S} and \ref{eq: explicitIon}, 
\begin{widetext}
    \begin{equation}\label{eq: explicit2S}
    \begin{split}
         \textbf{L}_{\textbf{2S}} = \int^\infty_{-\infty}  & \mathcal{N}\left(\mu, \sigma^2\right) dy \int^{2\pi}_0 d \varphi\int^\frac{\pi}{2}_0 \sin{2 \theta} 
       \\ & \times \int^\infty_0  (1-\lambda_{2S}) \frac{4 v^2 \pi^2  \beta_{ge}^2 I_0^2 w^2}{\sqrt{\pi} u^3 v_\perp^2} 
          \exp{ - \left(\frac{ 2\pi (\delta f + \Delta f_{AC} - \frac{v^2}{2c^2} f_0) w}{v_\perp}\right)^2 - \frac{D}{v_\perp\tau} - \frac{v^2}{u^2} - \frac{4 \rho^2}{w^2}} dv,
    \end{split}
\end{equation}
    \begin{equation}\label{eq: explicitIon}
    \begin{split}
         &\textbf{L}_{\textbf{ion}} =  \int^\infty_{-\infty} \mathcal{N}\left(\mu, \sigma^2\right)  dy \int^{2\pi}_0 d \varphi\int^\frac{\pi}{2}_0 \sin{2 \theta} 
       \\ & \times \int^\infty_0 (1-\lambda_i)
       \frac{1.4132 \pi^3\beta_i \beta^2_{ge} I_0^3 v^2 w^2}{\sqrt{\pi} u^3 v_\perp^2}
    \exp{-\frac{6\rho^2}{w^2} - \frac{D}{\tau v_\perp} -\frac{ v^2}{u^2}} V \left(\delta f + \Delta f_{AC} - \frac{v^2}{2c^2} f_0;1.1132 \eta ;  0.4535 \eta\right) dv.
    \end{split}
\end{equation}
\end{widetext}
 $\mathcal{N}\left(\mu, \sigma^2\right)$ is the Gaussian distribution of the positron beam in the $y$ direction, displayed in Fig.\ref{fig:laser_transit}, with center point $\mu$ and standard deviation $\sigma$. Parameters $\varphi$ and $\theta$ are the polar and azimuthal angles relative to a $z$ axis which points outwards from the target surface. The ionisation and two photon transition matrix elements, computed in \cite{Haas2006} are given by $\beta_i$ and $\beta_{ge}$ respectively. The variable $D$ is distance from target to laser, whereas $\rho$ and $v_\perp$ impact parameter and the velocity component perpendicular to the laser respectively, which contains an implicit dependence on $\theta$, $\varphi$, \& $y$ (see Fig. \ref{fig:laser_transit}). 

\begin{figure}[h!]
\includegraphics[width=\columnwidth]{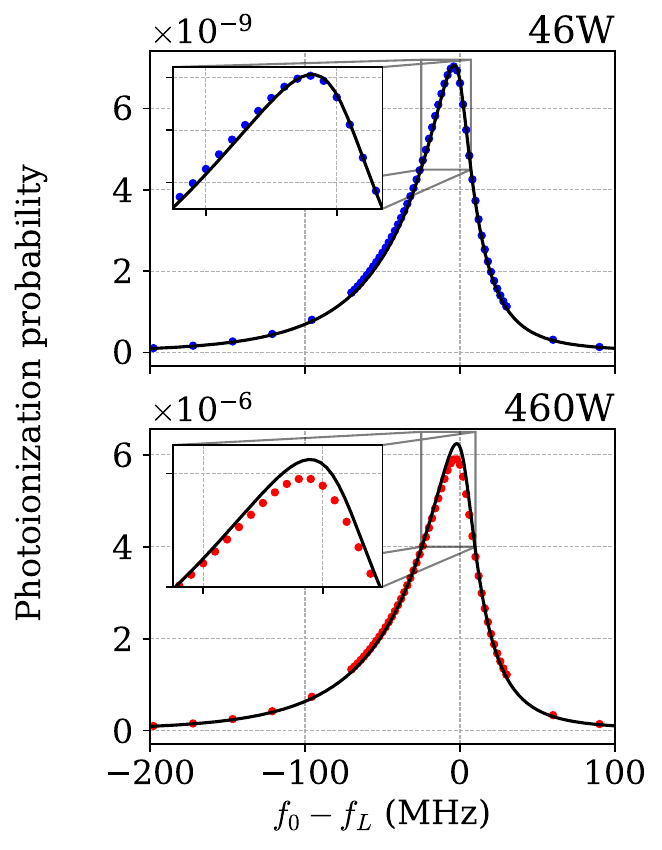} 
   \caption[] {The data points display the simulated photoionization spectra of Ps given the experimental conditions (see Tab. \ref{tab:exp_parameters}) while the solid lines plot $\textbf{L}_{\textbf{ion}}$ for two indicated powers: 46\,W and 460\,W,  the latter of these is the nominal power used to collect the experimental data. The inset displays the spectra near peak, highlighting the slight overestimate due to perturbation theory.
\label{fig:Lion}}
\end{figure}

These lineshape functions incorporate all the key physical effects present in our experimental setup.  Specifically, $\textbf{L}_{\textbf{ion}}$ is the expected lineshape from our experiment, given that our detection technique depends on photoionization from the 2S state by an additional 486\,nm photon within the same laser. A comparison between $\textbf{L}_{\textbf{ion}}$ and the Monte Carlo simulation of the full experiment is displayed for 46\,W and 460\,W intra-cavity power in Fig. \ref{fig:Lion}.

Since $\textbf{L}_{\textbf{ion}}$ is derived via perturbation theory, it is expected to become less accurate when the probability excitation to the 2S state or photoionization becomes significant, such as in the case of slow-moving atoms when the laser frequency is near resonance. This is evident for 460\,W (see Fig. \ref{fig:Lion}) near the peak of the lineshape, where $\textbf{L}_{\textbf{ion}}$ slightly overestimates the photoionization probability compared to the exact treatment; this difference is not as significant for the 46\,W spectrum.
\begin{figure}[h!]
\includegraphics[width=\columnwidth]{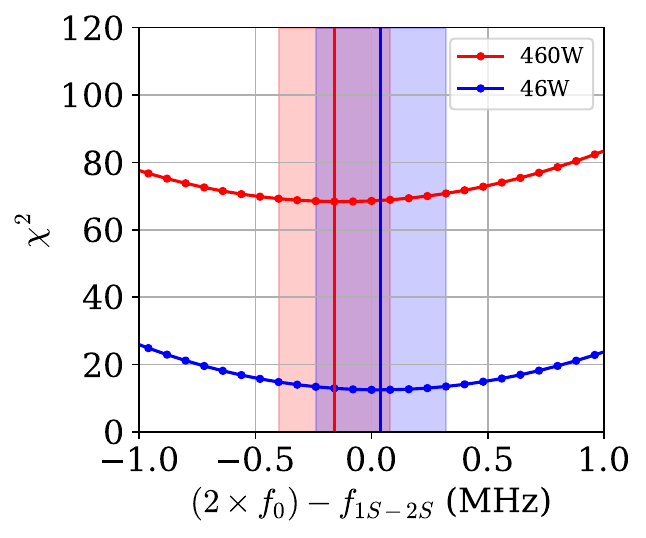} 
   \caption[] {Obtained $\chi^2$ as a function of $f_0$ (see text for more details).
\label{fig:chi_sq_sim}}
\end{figure}

To determine the effectiveness of $\textbf{L}_{\textbf{ion}}$ in extracting the transition frequency, we compare to the simulation for different values of the transition frequency $f_0$. By plotting the $\chi^2$ as a function of $f_0$ (see Fig. \ref{fig:chi_sq_sim}), the minimum provides the optimum fitted value, and the value one unit above the minimum corresponds to the uncertainty. For 46\,W and 460\,W the values of $2 \times f_0 - f_{1S-2S}$ is consistent with zero, with values 0.04$\pm$0.28\,MHz and 0.17$\pm$0.25\,MHz respectively. This indicates that the lineshape function is consistent with the simulation and can be used to extract the transition frequency for the given experimental conditions with an uncertainty of approximately 0.3\,MHz.

\section{\label{sec:results}Results and Analysis}

\subsection{Signal events and background suppression}

\begin{figure*}
\includegraphics[width=0.92\textwidth]{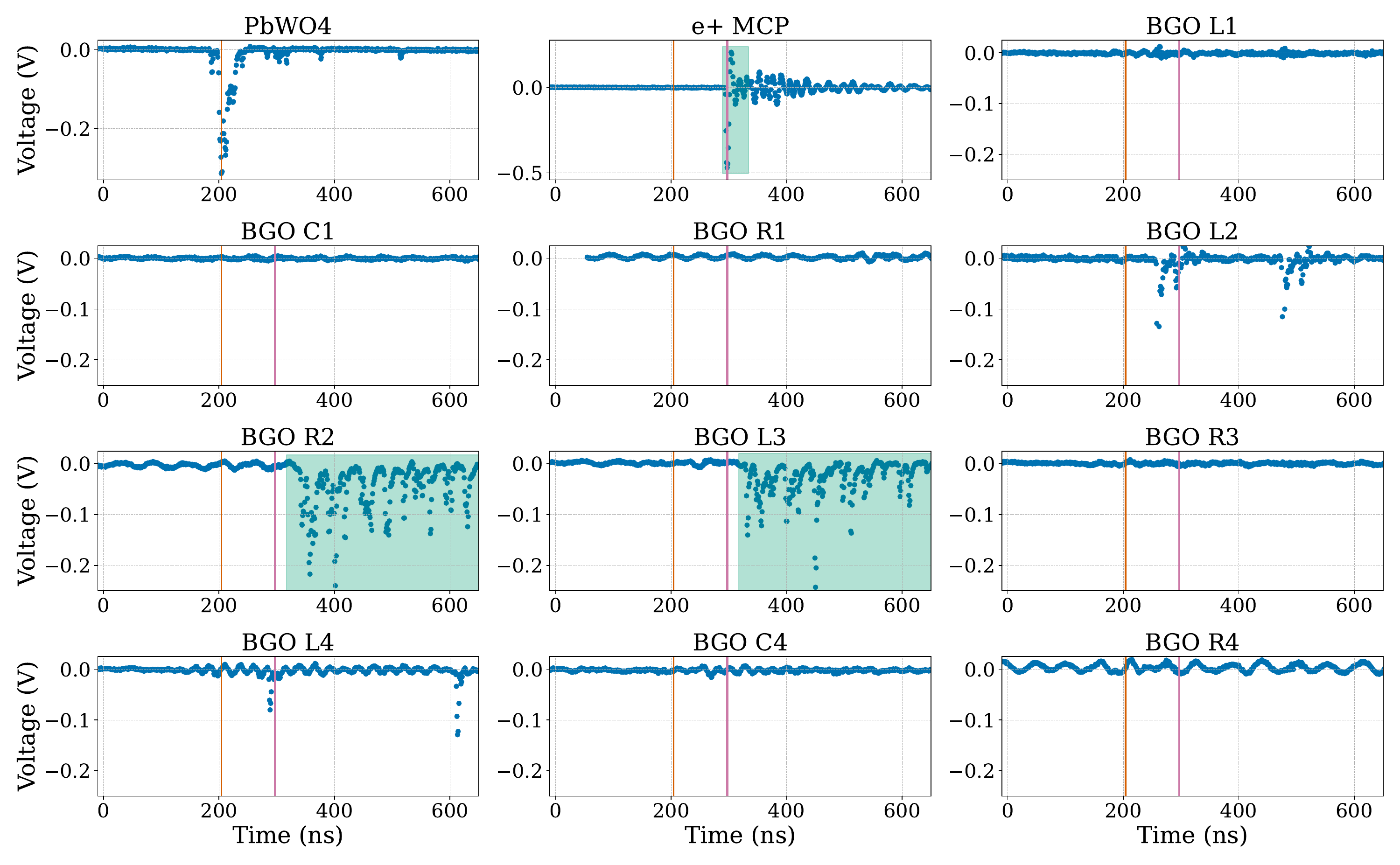}
   \caption[Detection of a signal event.]{Detection of a signal event: the positron pulse is flagged by the lead tungstate scintillator and its signal peak is represented by the orange vertical line. In the MCP channel, a count is recorded within the blue-green shaded region, corresponding to the \SIrange{85}{135}{\nano\second} time window post signal peak in the lead tungstate. Its time is designated by the vertical purple line. Shortly after, two \SI{511}{\kilo \electronvolt} $\gamma$ rays are detected in coincidence in BGOs R2 and L3 with their characteristic signature (blue-green shaded region). BGO L2 and L4 are respectively on top and below L3; as can be seen, they have a nonzero signal, which can be explained by light cross-talk.}
\label{fig:single_event}
\end{figure*}

A total of 1124 events satisfy the selection criteria of Section \ref{subsec:DAQ_analysis} after analysis of $\sim$12 days of data taking. Such an event satisfying the triple coincidence detection scheme is shown in Fig. \ref{fig:single_event}. The BGO combinations able to reconstruct a back-to-back annihilation are the pairs R1-L4, R2-L3, R3-L2, R4-L1 and C1-C4. From the measured lineshape (see Fig.~\ref{fig:lineshape_fit}), the signal-to-noise ratio (peak relative to baseline) exceeds 20, demonstrating the effectiveness of the detection scheme in suppressing background.

\subsection{Simulated Lineshape fitting}\label{subsec:MClineshapefitting}

\begin{figure*}
    \centering
    \includegraphics[width=0.6\textwidth]{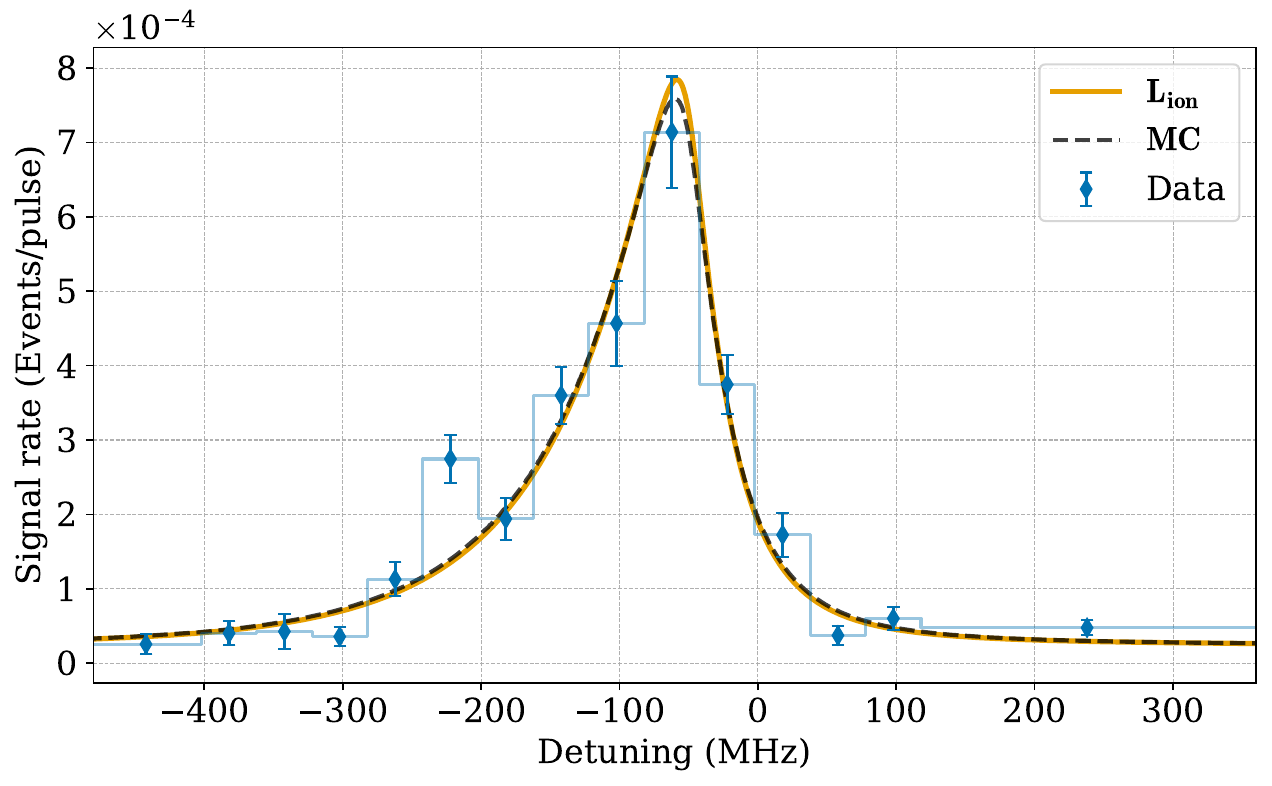}
    \caption{Measured spectrum of the $\text{1}^\text{3}\text{S}_\text{1} \to \text{2}^\text{3}\text{S}_\text{1}$ transition,  fitted with the semi-analytical lineshape $\mathbf{L_{ion}}$ and the Monte Carlo simulation $\mathbf{MC}$.}
    \label{fig:lineshape_fit}
\end{figure*}

High-statistics simulations (see Sec.~\ref{sec: MC Simulation} for more details) of a constrained parameter space are generated to fit the data. The simulations are then binned and fitted to the data for different values of the transition frequency $f_0$, where a $\chi^2$ test is performed. Two free parameters are used in the fit: a global amplitude to scale the simulation by the efficiency of the different experimental processes (positron-to-positronium conversion, guiding system efficiency, MCP quantum efficiency, back-to-back detection efficiency, and the raw number of positrons per pulse) and a global offset to account for the background. An artificial frequency offset was introduced during the analysis and removed only after the analysis was finalized in order to unblind the results. The minimum of the obtained $\chi^2$ distribution as a function of $f_0$ corresponds to the best-fit shift of the experimental value from the theoretical prediction. From this fitting procedure we extract a frequency shift of $-54.9\pm 2.4$\,MHz from the resonance, the corresponding fit to the spectrum is displayed in Fig. \ref{fig:lineshape_fit}.

The best-fit gives a background level of $2(1)\cdot 10^{-5}$ per positron pulse and an amplitude of $146(10)$, to be compared with the expected amplitude of $253(100)$ from Table~\ref{tab:detection_efficiency}.

\subsection{Analytical Lineshape fitting}\label{subsec:Analyticallineshapefitting}
$\textbf{L}_{\textbf{ion}}$ can also be fit to the data to extract the transition frequency. As with fitting the simulation, the positron pulse size, mean positron temperature, laser-target distance, and laser power were all fixed to the nominal values given in section \ref{subsec:MClineshapefitting}. We apply an equivalent blinded fitting procedure as in section \ref{subsec:MClineshapefitting},  the lineshape was then binned to the equivalent values as the experimental data, and fit to the data for different transition frequencies, $f_0$. The data could then be fit with the function
\begin{equation}
   A\times\textbf{L}_{\textbf{ion}}(f_0) + c
\end{equation}
where $A$ is a general scaling parameter, and $c$ is an offset to account for background. From this fitting procedure we extract a frequency shift of $-54.4\pm 2.8$\,MHz (see Fig. \ref{fig:lineshape_fit}) in very good agreement with the value obtained using the full MC simulation.

\subsection{Systematic uncertainties}\label{sec:systematics}

As anticipated in Sec.~\ref{sec:MCsimulation}, the systematic uncertainties associated with the experimental parameters are evaluated using dedicated simulations.

The intense intracavity laser field used to drive the two-photon 1S–2S transition induces a shift in the energy levels of the positronium atom via the AC Stark effect. This shift is proportional to the intensity of the electric field, and therefore scales linearly with the intracavity power. To quantify this effect, MC simulations were performed at multiple intracavity powers. The uncertainty on this correction arises from the precision of the power calibration, and shot-to-shot power fluctuations during data acquisition. A conservative estimate yields a total uncertainty of 0.4 MHz.

The two-photon excitation probability depends on the overlap between the spatial distribution of the positronium and the laser mode in the enhancement cavity. An offset between the laser waist and the mean formation point of the atoms results in asymmetries in the excitation profile and a shift of the observed line center. This systematic effect was evaluated using simulations and sensitivity studies of the lineshape to shifts in the laser position, which lead to a conservative uncertainty of 1 MHz.

The uncertainty on the velocity distribution translates into a systematic contribution via the second-order Doppler shift at the level of 0.3 MHz.

The electric field used to guide the photoionized positrons to the detection region induces a DC Stark shift of the transition, which is evaluated to be 66 kHz (see derivation in Appendix~\ref{sec:appendix_DCStarkShift}). The uncertainty on the electric field in the excitation region is included as a separate contribution of 4 kHz in the systematic budget.

The main source of uncertainty in our measurement is related to frequency metrology relying on a wavemeter calibrated with a rubidium standard. The details are given in Appendix~\ref{sec:appendix_frequency_metrology}.

All systematic uncertainties are summarized in Table~\ref{tab:error_budget} and are added in quadrature to the statistical uncertainty to obtain the total uncertainty.

\subsection{Result}\label{sec:Result}

\begin{figure}
    \centering
    \includegraphics[width=\linewidth]{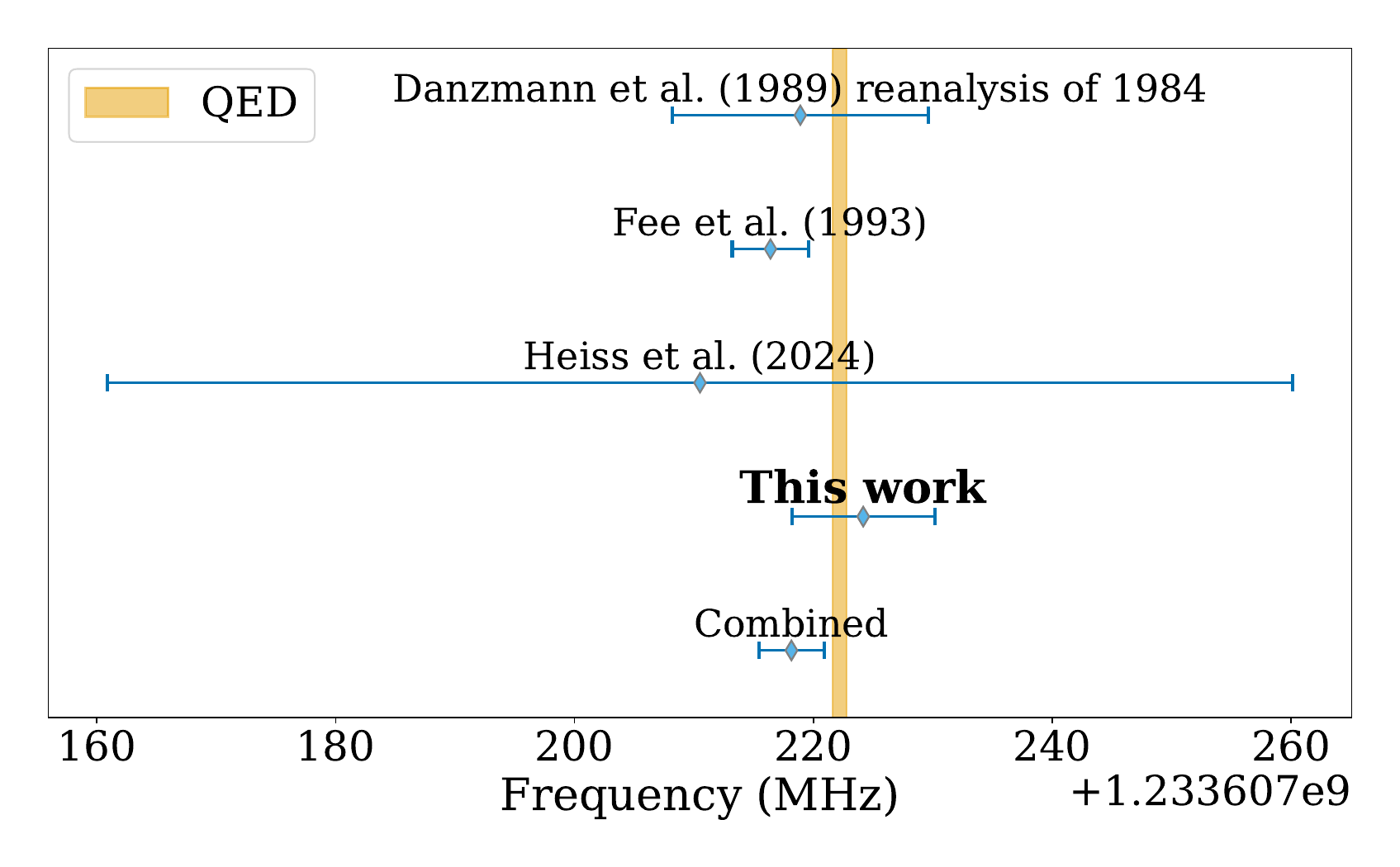}
    \caption{Compilation of positronium $\text{1}^\text{3}\text{S}_\text{1} \to \text{2}^\text{3}\text{S}_\text{1}$ measurements and theoretical prediction of QED at the order $\mathcal{O}(\alpha^7\ln^2(1/\alpha))$.}
    \label{fig:ps1s2s_measurements}
\end{figure}

The transition is thus measured to be at $\numprint{1233607224.1}(6.0)\,\text{MHz}$ (4.9 ppb) and agrees with the theoretical prediction of $\numprint{1233607222.12}(58)\,\text{MHz}$ and the previous precision measurements of this transition as shown in Fig. \ref{fig:ps1s2s_measurements}. Our measurement is dominated by the laser frequency metrology error, followed by statistics and accuracy in the laser position. The weighted average of all the available measurements amounts to $\numprint{1233607218.1}(2.7)\,\text{MHz}$, resulting in a reduced tension of $1.4\,\sigma$ with the most recent QED developments \cite{ADKINS20221}.\\

\begin{table}[!htbp]
\centering
\begin{tabular}{>{\bfseries}l c r}
\hline\hline
\textbf{} & \textbf{Shift (MHz)} & \textbf{$\sigma$ (MHz)} \\
\hline
Fitting MC/$\textbf{L}_{\textbf{ion}}$ & -54.9/-54.4 & 2.4/2.8\\
\hline
Frequency Correction & -- & 3.6 \\
Wavemeter Systematic & 56.8 & 2.8 \\
Frequency measurement & -- & 2.0 \\
Rubidium reference & -- & 2.0\\
\hline
Laser Position & -- & 1.0 \\
AC Stark & (-4.9) & 0.4 \\
Second-order Doppler & (9.4) & 0.3 \\
DC Stark & 0.066 &  0.004 \\
\hline
\textbf{Deviation to theory} & 2.0 & 6.0 \\
\hline\hline
\end{tabular}
\caption[Shift and uncertainty estimates.]{Shift and uncertainty estimates. Shift values in parentheses are already included in the fitting shift.}
\label{tab:error_budget}
\end{table}

\section{Conclusion and outlook}

We have presented a precision measurement of the $\text{1}^\text{3}\text{S}_\text{1} \to \text{2}^\text{3}\text{S}_\text{1}$ positronium interval using continuous-wave two-photon laser spectroscopy. Our result is consistent with previous continuous-wave measurements and with the most recent QED predictions. The experiment also demonstrated the efficiency of the triple-coincidence detection scheme in suppressing background and achieving a signal-to-noise ratio greater than 20.

In addition, we developed and validated a semi-analytical lineshape model that captures physical effects such as the initial spatial and velocity distributions, second-order Doppler shifts, AC Stark shifts, finite lifetimes, and photoionization. The model accurately reproduces the experimental data and and provides an efficient alternative to Monte Carlo simulations. The transition frequency extracted using this model is in very good agreement with the value obtained from the MC-based method.

With improved laser frequency metrology using a frequency comb, higher excitation efficiency from operation in a quasi-CW regime that allows increased laser power without cavity degradation \cite{Zhadnov:2023vcr}, and 20 times more statistics by employing a new positron source, the scheme presented here will enable sub-ppb precision. One promising direction to reach an uncertainty at the few-ppt level is the implementation of the so-called Ramsey–Doppler spectroscopy method \cite{Javary2025}, which offers a route to reduce by a factor 100 the experimental linewidth and suppress systematic effects such as the AC stark shift. Alternatively, the application of laser cooling techniques to Ps \cite{Shu:2023fgm, AEgIS:2023lpw} could also be used to reduce systematic uncertainties, such as transit-time broadening and Doppler shifts, by lowering the kinetic energy of the atoms, thereby providing an independent path toward enhanced spectral resolution. Such advances would enable a stringent test of bound-state QED and open new opportunities to search for physics beyond the Standard Model.

\section*{Acknowledgements}

We are in debt with A. Antognini, K. Kirch and F. Merkt for their continuous and essential support on this project. We are grateful to T. Esslinger and T. Donner for allowing us to use their Rubidium reference. We acknowledge D. Cooke, A. Golovizin, P. Gherardi, M. Heiss and G. Wichmann for their substantial contributions to the development of the apparatus which provided the basis for these results. 
This work was supported by the ERC consolidator grant 818053-Mu-MASS, the Swiss National Science Foundation under the grants 197346, 219485 and NIST PMG grant 60NANB24D321.

\appendix

\section{\label{sec:appendix-lineshape} Semi-analytical lineshape derivation}

Monte Carlo simulations which evaluate the Optical Bloch Equations (OBE) can fully model the experimental conditions. However, these simulations lack intuitiveness and can often be perceived as a "black box" approach, which is computationally demanding, especially when numerous simulated spectra are necessary to fit the experimental data. To reduce dependence on simulations and provide a clearer mathematical description of the $\text{1}^\text{3}\text{S}_\text{1} \to \text{2}^\text{3}\text{S}_\text{1}$ positronium spectrum, we have developed a semi-analytical lineshape model based on second-order perturbation theory.

\subsection{Monoenergetic beam}\label{sec:appendix_monoenergetic}
To start with, we take the assumption that Ps traverses the laser with a constant velocity perpendicular to the laser axis, $v_\perp$, passing its point of closest approach at time $t=0$. As such, the radial coordinate of the atom in the $x-y$ plane with respect to the laser axis varies with $t$ as $R^2(t) = \rho^2 + v^2_{\perp}t^2$, where $\rho$ is the impact parameter with the laser axis. The positronium trajectory as it traverses the laser beam is visualised in Fig. \ref{fig:laser_transit}. 
For an atom on a straight trajectory through a Gaussian beam with width of $w$ and a peak intensity of $I_0$, the transition amplitude, $C_{2S}$ can be approximated by \cite{Biraben1979}
\begin{align}\label{eq: C_2S_t}
\begin{split}
        C_{2S}(t) & =  -i \cdot 
        2 \pi \beta_{ge} I_0 e^{-2 \rho^2/w^2} \\ &\int^t_{-\infty} dt^{\prime}  e^{-2 v^2_{\perp}{t^{\prime}}^2/w^2}    e^{i (\omega_{0} - 2 \omega_L)t^{\prime }},
\end{split}
\end{align}
where $\omega_{0}/2\pi$ is the resonant frequency of the transition, $\omega_L/2\pi$ is the laser frequency, and $\beta_{ge}$ is time-independent
two-photon transition matrix element. The value of $\beta_{ge}$ as calculated in \cite{Haas2006}
 for hydrogen has been scaled by a factor of 8 to account for the different reduced mass of positronium. The integral can be evaluated between $(-\infty,\infty)$ to determine the change of states during one complete pass of the laser. This gives
\begin{align}\label{eq: C2_const_v}
    C_{2S} & =  -i \cdot 2\pi\beta_{ge} I_0
     e^{-2 \rho^2/w^2} \frac{w\sqrt{\pi}}{v\sqrt{2}}e^{-\frac{(\omega_{0} - 2 \omega_L)^2 w^2}{8v_{\perp}^2}}.
\end{align}
The magnitude of this amplitude provides the probability of being excited to the $2S$ state, resulting in a Gaussian lineshape,  given by
\begin{align}
\begin{split}
    \label{eq: P_const_v}
    P_{2S}  & = |C_{2S}|^2 =  2\pi^3 \beta_{ge}^2 I_0^2 e^{-4 \rho^2/w^2} \frac{w^2}{v_\perp^2}e^{ - \left(\frac{ 2\pi (f_0 - f_L) w}{v_\perp}\right)^2} ,
\end{split}
\end{align}
where we have defined the two photon resonant frequency $\omega_{0} = 2 \cdot 2 \pi f_0$ and converted back  to regular frequency for the laser $\omega_{L} = 2 \pi f_L$.

\subsection{Ionisation within spectroscopy laser}
The detection mechanism for the experimental setup described in section \ref{sec:experimental} relies on photoionization from the 2S state by an additional 486nm photon whilst traversing the spectroscopy laser. The lineshape detected via photoionization differs from the Gaussian lineshape for a monoenergetic beam of Ps described in equation \ref{eq: P_const_v}, as photoionization reduces the lifetime of the 2S state, introducing an additional source of broadening. To account for the reduced lifetime, we evaluate the integral in equation \ref{eq: C_2S_t}, to time $t$ to find the time dependence of $C_{2S}$,
\begin{align}
\begin{split}\label{eq: C2S_t_dep}
    C_{2 S}^{(2)}(t) = &-i\pi \beta_{ge} I_0
    e^{-2 \rho^2/w^2} \frac{\sqrt{\pi} w}{\sqrt{2} v_\perp} e^{-\frac{(\omega_{0} - 2 \omega_L)^2 w^2}{8 v_\perp^2}}\\&\left[1+\operatorname{erf}\left(\frac{\sqrt{2} v_\perp t}{w}-i \frac{(\omega_{0} - 2 \omega_L) w}{2 \sqrt{2} v_\perp}\right)\right].
\end{split}
\end{align}

In the limit where $t \rightarrow \infty$ the value of the square bracket converges to two, and equation \ref{eq: C2_const_v} is recovered. However, outside of this limit, the complex term within the error function has a frequency dependence, which introduces a broadening. The probability of an atom being photoionized during one complete pass of the laser can be expressed as
\begin{equation}\label{eq: P_ion_int}
    P_{i}=\int_{-\infty}^{\infty}\gamma_{i}(t) |C_{2S}(t)|^2 dt,
\end{equation}
where $\gamma_{i}(t)$ is the ionisation rate from the 2S state in the same laser. In the above expression we have taken the perturbative assumption that photoionization does not alter the population of the 2S state during a single pass of the laser. The photoionization rate is given by 
\begin{equation}\label{eq: gam_ion}
    \gamma_{i}(t) = 2 \pi \beta_{i} I_0 \exp(-\frac{2\rho^2}{w^2}) \exp(-\frac{2 v_\perp^2 t^2}{w^2}), 
\end{equation}
where $\beta_i$ represents the ionisation coefficient as defined in \cite{Haas2006}, again, scaled by a factor of 8 to account for different reduced mass. By substituting equations \ref{eq: C2S_t_dep} and \ref{eq: gam_ion} into equation \ref{eq: P_ion_int}, and after some simplification, it is possible to obtain
\begin{equation}\label{eq:P_ion_int_2}
\begin{split}
    P_{i} =  \pi^4\beta_i \beta^2_{ge} I_0^3 \frac{w^2}{v_\perp^2}&
    e^{-\frac{6\rho^2}{w^2}} e^{-\frac{\delta f^2}{2 \eta^2}} \\ \times \int_{-\infty}^{\infty}  e^{-(4 \pi \eta t)^2} &  \left|1+\operatorname{erf}\left(4 \pi \eta t -i \frac{\delta f}{2 \eta}\right)\right|^2 dt,
\end{split}
\end{equation}
where $\delta f = f_0 - f_L = (\omega_0 - 2\omega_L)/4 \pi $, and we define $\eta = \frac{v_\perp}{2 \sqrt{2} \pi w}$.  Whilst this integral contains complex values internally, its total value is real and positive due to the modulus squared. Whereas it is possible to evaluate this function numerically, this can become computationally expensive, particularly when nested within other numerically evaluated integrals detailed in the following subsections. To circumvent this, we find that equation \ref{eq:P_ion_int_2} can be conveniently approximated as 
\begin{equation}\label{eq:P_ion_Vgt}
\begin{split}
    P_{i} \approx  0.7066 \pi^4\beta_i \beta^2_{ge} & I_0^3 \frac{w^2}{v_\perp^2}
    e^{-\frac{6\rho^2}{w^2}}\\ &V (\delta f;1.1132 \eta ;  0.4535 \eta)
\end{split}
\end{equation}
where $V( \delta f; \sigma; \gamma)$ is a Voigt function characterised by Gaussian and Lorentzian widths of $\sigma$ and $\gamma$, respectively. The values of $\sigma=1.1132 \eta $ and $\gamma=0.4535 \eta $ are found to provide a good fit to the integral in equation \ref{eq:P_ion_int_2}, which remains general over a wide range of $\eta$.
By applying this approximation instead of directly evaluating the integral in \ref{eq:P_ion_int_2}, we can take advantage of highly optimised algorithms to generate Voigt functions which exist in most modern programming languages.

\subsection{Depletion of 2S state}
So far, we have taken the perturbative assumption that the 1S state population is not depleted by the excitation, and the 2S state population is not depleted by photo\-ionisation. The former of these assumptions remains a good approximation even at the high laser powers required for positronium experiments. However, the latter assumption, that the 2S state is not depleted, becomes a less accurate approximation near the resonance for higher laser powers, or lower atomic velocities. For example, for our nominal experimental parameters (taking an average velocity of 95\,km/s) we find the photo\-ionisation probability given by Eq.~\ref{eq:P_ion_Vgt} is around 10\% of the $\text{1}^\text{3}\text{S}_\text{1} \to \text{2}^\text{3}\text{S}_\text{1}$ excitation probability given by Eq.~\ref{eq: P_const_v}. To account for these depletion effects, we can add extra perturbative terms to Eq.~\ref{eq: P_const_v}.

\begin{equation}\label{eq:corrected_P2S int}
    P_{2S}^* \;=\; P_{2S} \;-\; \underbrace{\int_{-\infty}^{\infty} \gamma_i(t)\,\big|C_{2S}(t)\big|^2\,dt}_{\Delta P_{2S}}\,.
\end{equation}
Here \(P_{2S}^*\) is the probability of being in the 2S state with a first–order correction for depletion due to ionisation, \(\Delta P_{2S}\). As depletion due to photo\-ionisation is a perturbative correction that is most relevant near resonance, we substitute Eq.~\ref{eq: C2S_t_dep} and approximate the imaginary term in the error function by zero. This yields
\begin{equation}
\begin{split}
\Delta P_{2S} \;\approx\;&\; 2\pi\,\beta_{i}\,I_0\,e^{-2\rho^2/w^2}\;\frac{P_{2S}}{4} \\
&\times \int_{-\infty}^{\infty} \exp\!\Big(-\frac{2 v_\perp^{2} t^{2}}{w^{2}}\Big)\,\Big[1+\operatorname{erf}\!\Big(\frac{\sqrt{2}\,v_\perp t}{w}\Big)\Big]^2 dt,
\end{split}
\end{equation}
where the integral can be evaluated directly to give
\begin{equation}
\Delta P_{2S}
\;=\; \pi\,\beta_{i}\,I_0\,e^{-2\rho^2/w^2}\;\frac{w\sqrt{2\pi}}{3 v_{\perp}}\;P_{2S}
\;\equiv\; \lambda_{2S}\,P_{2S}.
\end{equation}
Hence,
\begin{equation}
    P^*_{2S} \;=\;  (1 -\lambda_{2S})\, P_{2S}\,.
\end{equation}

An equivalent approach applies to the ionisation probability \(P_i\): add a first–order perturbative term to account for depletion of the 2S state during ionisation,
\begin{equation}\label{eq: integral correction Pi}
    P_{i}^* \;\approx\;  P_i \;-\;  \underbrace{\int_{-\infty}^{\infty} \gamma_i(\tau)\!\int_{-\infty}^{\tau} \gamma_i(s)\,\big|C_{2S}(s)\big|^2 ds\,d\tau}_{\Delta P_{i}}\,,
\end{equation}
where \(P_i=\int\gamma_i(t)\,|C_{2S}(t)|^2 dt\) is the no–depletion ionisation probability. By Tonelli’s theorem (non–negative integrand) we may reorder the integrals in \(\Delta P_{i}\),
\begin{equation}
\Delta P_{i} \;=\;\int_{-\infty}^{\infty}\!\Big(\int_{\tau=s}^{\infty} \gamma_i(\tau)\, d \tau\Big)\, \gamma_i(s)\,\big|C_{2 S}(s)\big|^2\, d s,
\end{equation}
and with \(\gamma_i(t)=\Gamma\,e^{-\alpha t^2}\) (\(\Gamma=2\pi\beta_{i} I_0 e^{-2\rho^2/w^2}\), \(\alpha=2v^{2}/w^{2}\)) this becomes
\begin{equation}
\begin{split}
\Delta P_{i}
= \frac{\Gamma^2 \sqrt{\pi}}{2 \sqrt{\alpha}} \;\frac{P_{2 S}}{4} 
&\int_{-\infty}^{\infty}  e^{-\alpha s^{2}}\,
\operatorname{erfc}(\sqrt{\alpha}\, s)\,\\
&\times \Big|1+\operatorname{erf}\!\Big(\sqrt{\alpha}\,s - i \frac{\delta f}{2 \eta}\Big)\Big|^{2}  d s.
\end{split}
\end{equation}
Assuming the correction is most relevant near resonance, we set \(\delta f \approx 0\) and evaluate the remaining integral, obtaining
\begin{equation}
\Delta P_i \;=\; \frac{\pi}{12}\,\frac{\Gamma^{2}}{\alpha}\;P_{2S}.
\end{equation}
On resonance we may relate \(P_{2S}\) to \(P_i\) via
\begin{equation}
P_i=\int_{-\infty}^{\infty} \gamma_i(t)\,\big|C_{2 S}^{(2)}(t)\big|^2\, dt
= \frac{\sqrt{\pi}}{3}\,\frac{\Gamma}{\sqrt{\alpha}}\;P_{2S},
\end{equation}
This allows us to define the entire correction conveniently in terms of a scaling on $P_i$.
As such, we can express equation \ref{eq: integral correction Pi} as 
\begin{equation}
    P_i^* = (1-\lambda_i)P_i.
\end{equation}
where 
\begin{equation}
\Delta P_i=\frac{\Gamma \sqrt{\pi}}{4 \sqrt{\alpha}} P_{i} = \lambda_i P_i.
\end{equation}

\subsection{Second-order Doppler effect \& self annihilation}\label{sec: DopMod}
\begin{figure}[t]
\includegraphics[width=0.95\columnwidth]{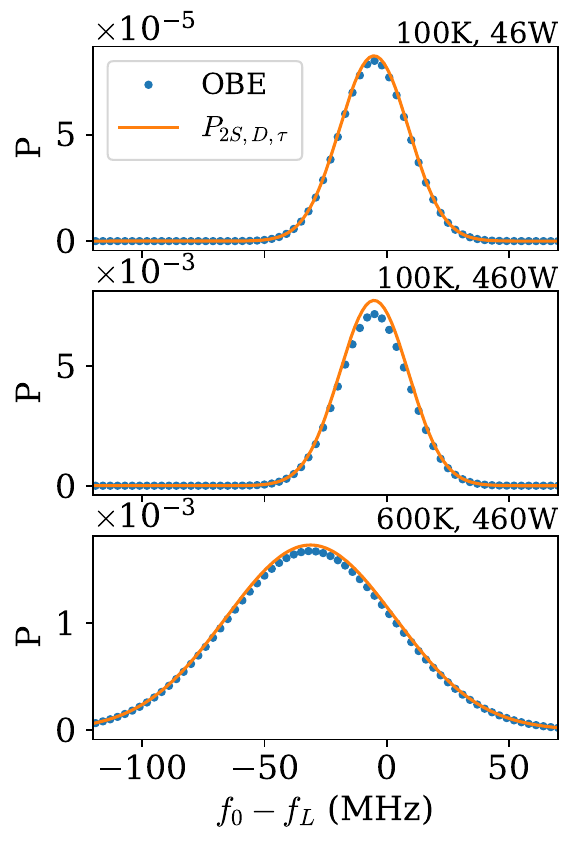} 
   \caption[] {Simulated spectra of different powers and velocities alongside $P_{2S, D, \tau}$. Note that the velocity is given in terms of units of temperature through the conversion $T = \frac{1}{2 k_B}m v^2$. Remaining variables are set to $w_0$= 310\,$\mu$m, $D$ = 2\,mm, $\rho=0$. 
\label{fig:P2Sdt}}
\end{figure}
\begin{figure}[t]
\includegraphics[width=\columnwidth]{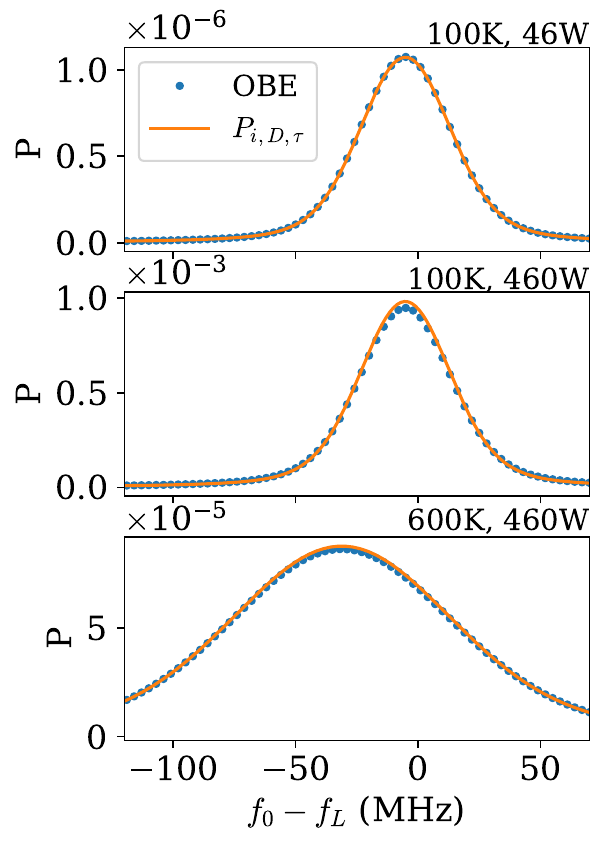} 
   \caption[] {Simulated spectra of different powers and velocities alongside $P_{i, D, \tau}$. Note that the velocity is given in terms of units of temperature through the conversion $T = \frac{1}{2 k_B}m v^2$. Remaining variables are set to $w_0$= 310\,$\mu$m, $D$ = 2\,mm, $\rho=0$. 
\label{fig:Pidt}}
\end{figure}
At this point, we can incorporate the modifications due to the second-order Doppler effect, and the finite lifetime of o-Ps in the ground state due to self-annihilation.  Due to the light mass of Ps, it exhibits high velocities even at relatively low energies - meaning that the second-order Doppler effect is more relevant than it is in typical atoms. In the lab frame of reference, the Ps will exhibit a modified one-photon transition frequency, $f_0$, due to the second-order Doppler effect, which depends on the magnitude of its velocity in accordance to
\begin{equation}
f^\prime_{0} = f_0\sqrt{1-\frac{v^2}{c^2}}\approx f_0 -\frac{v^2}{2c^2} f_0,
\end{equation}
where $c$ is the speed of light. Bear in mind this expression differs by a factor of 1/2 from similar expressions in \cite{Haas2006} as we are using the one photon transition frequency $f_0$ as defined in section \ref{sec:appendix_monoenergetic}. This can be straightforwardly integrated into equation \ref{eq: P_const_v} and \ref{eq:P_ion_Vgt} by substituting $f_0 \rightarrow f^\prime_{0}$, giving
\begin{align}
\begin{split}
    \label{eq: P_const_v_dop}
    P_{2S, D}  =  & (1-\lambda_{2S}) 2\pi^3 \beta_{ge}^2 I_0^2  e^{-4 \rho^2/w^2} \\ \times \frac{w^2}{v_\perp^2}
    & \exp{ - \left(\frac{ 2\pi (f_0 - f_L - \frac{v^2}{2c^2} f_0) w}{v_\perp}\right)^2}.
\end{split}
\end{align}
\begin{equation}\label{eq:P_ion_Vgt_dop0}
\begin{split}
    P_{i, D} =  0.7066 &  (1-\lambda_{i}) \pi^4\beta_i \beta^2_{ge} I_0^3 \frac{w^2}{v_\perp^2}
    \exp{-\frac{6\rho^2}{w^2}}\\ &V (\delta f - \frac{v^2}{2c^2} f_0;1.1132 \eta ;  0.4535 \eta) .
\end{split}
\end{equation}

The finite lifetime, $\tau$, of the Ps due to self-annihilation can be integrated into the expression. The probability of a Ps surviving whilst traveling the distance, $D$, between the target and the laser, displayed in Fig. \ref{fig:laser_transit}, is given by $\exp{(-\frac{D}{v_\perp\tau})}$. This introduces a straightforward multiplicative factor in front of \ref{eq: P_const_v_dop} and \ref{eq:P_ion_Vgt_dop0} to give
\begin{align}
\begin{split}
    \label{eq: P_const_v_dop_fin}
    P_{2S, D, \tau}&  =  (1-\lambda_{2S}) 2\pi^3 \beta_{ge}^2 I_0^2  e^{-4 \rho^2/w^2} \\ \frac{w^2}{v_\perp^2}
    &  \exp{ - \left(\frac{ 2\pi (f_0 - f_L - \frac{v^2}{2c^2} f_0) w}{v_\perp}\right)^2 - \frac{D}{v_\perp\tau}} ,
\end{split}
\end{align}
\begin{equation}\label{eq:P_ion_Vgt_dop}
\begin{split}
    P_{i, D, \tau} =  0.7066 & (1-\lambda_{i}) \pi^4\beta_i \beta^2_{ge} I_0^3 \frac{w^2}{v_\perp^2}
    \exp{-\frac{6\rho^2}{w^2} - \frac{D}{\tau v_\perp}}\\ &V (\delta f - \frac{v^2}{2c^2} f_0;1.1132 \eta ;  0.4535 \eta) .
\end{split}
\end{equation}

Figures \ref{fig:P2Sdt} and \ref{fig:Pidt} display a comparison of the lineshape functions in equations \ref{eq: P_const_v_dop_fin} and \ref{eq:P_ion_Vgt_dop} with spectra computed by Monte Carlo simulations, which evaluate the optical Bloch equations for a monodirectional, monoenergetic positronium beam. Note that in this monodirectional, monoenergetic picture we set $v = v_\perp$, and that the AC Stark shift is not included in either the simulations or the lineshape at this point; it is addressed in later sections.  It is evident that at higher powers and lower temperatures, some of the perturbative assumptions of the lineshape become less effective. This results in the lineshape overestimating the probability of excitation and photoionization when the laser is close to resonance

\subsection{Thermal distribution}
The lineshape functions described above are derived only for a monoenergetic distribution of atoms. However, this is not the case in our experimental setup; instead, each incoming pulse of positrons on the target produces positronium with a spread of velocities. It has been observed that the magnitude of the velocity with which the Ps emitted from the target can be modelled by a 3D Maxwell-Boltzmann distribution \cite{PhysRevA.111.012810}:
\begin{equation}
\begin{split}
g(v) = \frac{4 v^2}{\sqrt{\pi} u^3}\exp \left(-\frac{ v^2}{u^2}\right) ,
\end{split}
\end{equation}
where $u=\sqrt{2 k_B T/m}$  denotes the characteristic velocity of the thermal distribution, which depends on the particle mass $m$ and the distribution temperature $T$, where $k_B$ is the Boltzmann constant. The direction of emission from the target, which dictates the individual velocity components of the Ps, is found to be well modelled by a $\cos{\theta}$ distribution \cite{Greenwood2002}. This results in velocities which are predominantly perpendicular to both the emittance surface and laser axis. For the time being, to simplify things we assume that Ps is emitted entirely perpendicularly to the surface, leading to $v_z \approx v_\perp \approx v$. The effects of the full $\cos{\theta}$ distribution on the lineshape discussed in  section \ref{sec: cosine}.

To derive a lineshape, $L$, corresponding to the spectra observed for a thermal ensemble of atoms, we integrate the fixed velocity lineshapes derived previously over the Maxwell-Boltzmann distribution
\begin{align}\label{eq: MB integral}
   L &= \int^\infty_0 g(v) P(v) dv,
\end{align}
where $P(v)$ can be any of the lineshape derived above. Unfortunately, once the finite lifetime term $\exp \left\{\left(-\frac{D}{v_{\perp} \tau}\right)\right\}$ is introduced, the integral can no longer be solved analytically. As such, these integrals must be solved numerically.

\subsection{Angular distributions, spatial distributions and the AC Stark shift}\label{sec: cosine}
In the above expressions we have assumed atoms are emitted perpendicularly to the surface of the target, entirely in the $\vec{z}$ direction. However, in reality, the emission of atoms occurs across a range of polar and azimuthal angles, $\theta$ and $\varphi$, which can be modelled as a cosine distribution \cite{Greenwood2002}. Furthermore, the positronium is not emitted from a point-like location on the target, instead they are formed within the areas which the positron beam impinges upon the surface of the target, which is modelled as a 2D Gaussian.  

We can define the velocity perpendicular to the laser axis, $v_\perp$, which parametrises the transit-time broadening and time-of-flight to the laser. If we define the laser axis in the $y$ direction, then
\begin{equation}
    v_\perp = \sqrt{v_x^2 + v_y^2} = v\sqrt{\cos^2\theta +\sin^2\theta|\cos^2\varphi|}.
\end{equation}

The impact parameter, $\rho$, can be defined in relation to the angle of emission, the point of formation of Ps in the dimension orthogonal to the laser axis on the target surface (illustrated as $y$ in Fig. \ref{fig:laser_transit}), and the separation between the target and the laser, denoted as $D$.
\begin{equation}
    \rho = D \sin\left( \tan^{-1}(\tan\theta \cos \varphi) + \frac{y}{D}\right)
\end{equation}
Additionally, employing the dressed-state approach presented in \cite{quasi_analytical}, the influence of $\rho$ on the AC-Stark effect of the $\text{1}^\text{3}\text{S}_\text{1} \to \text{2}^\text{3}\text{S}_\text{1}$ transition can be estimated as 
\begin{equation}
\Delta f_{AC}\left(\rho\right) = \frac{1}{2} \times \frac{\Delta \omega_{\mathrm{AC}}}{2 \pi}=\sqrt{\frac{2 \pi}{3}} \beta_{\mathrm{AC}} \pi I_0 e^{-\frac{2 \rho^2}{w^2}},
\end{equation}
where $\beta_{\mathrm{AC}}$ is the AC Stark coefficient derived in \cite{Haas2006}, scaled by a factor of 8. The effects of the AC stark effect can straightforwardly be incorporated into the lineshape by making the substitution
\begin{equation}
    f_0 \rightarrow f_0 + \Delta f_{AC}\left(\rho\right).
\end{equation}

With these new definitions, we can make substitutions for $v_\perp$, $\rho$ and $f_0$ in equations \ref{eq: P_const_v_dop_fin} and \ref{eq:P_ion_Vgt_dop}. To account for the distribution of emission angles and formation of the Ps emitted from the target, we numerically evaluate integrals over $\theta$ and $\phi$, weighted by a cosine probability distribution \cite{Greenwood2002}, and an additional integral over $y$ weighted by a Gaussian,
\begin{equation}\label{eq: quad_integral}
    \begin{split}
   \textbf{L} = \int^\infty_{-\infty} \mathcal{N}\left(\mu, \sigma^2\right) &dy \int^{2\pi}_0\frac{1}{2 \pi}\int^\frac{\pi}{2}_0 \sin{2 \theta} \\
   &\int^\infty_0 g(v) P(v, \phi, \theta, y) dv,
   \end{split}
\end{equation}
where $\mathcal{N}\left(\mu, \sigma^2\right)$ is a Gaussian distribution with a mean $\mu$, and standard deviation 
$\sigma$, derived from the positron beam profile. For clarity we state the entire expressions for the 2S and ionisation lineshape below in equations \ref{eq: explicit2SApp} and \ref{eq: explicitIonApp}, where $\rho$, $v_\perp$ \& $f_{AC}$ have implicit dependence on $\theta$, $\varphi$, \& $y$ as stated above,

\begin{widetext}
    \begin{equation}\label{eq: explicit2SApp}
    \begin{split}
         \textbf{L}_{\textbf{2S}} = \int^\infty_{-\infty}  & \mathcal{N}\left(\mu, \sigma^2\right) dy \int^{2\pi}_0 d \varphi\int^\frac{\pi}{2}_0 \sin{2 \theta} 
       \\ & \times \int^\infty_0  (1-\lambda_{2S}) \frac{4 v^2 \pi^2  \beta_{ge}^2 I_0^2 w^2}{\sqrt{\pi} u^3 v_\perp^2} 
          \exp{ - \left(\frac{ 2\pi (\delta f + \Delta f_{AC} - \frac{v^2}{2c^2} f_0) w}{v_\perp}\right)^2 - \frac{D}{v_\perp\tau} - \frac{v^2}{u^2} - \frac{4 \rho^2}{w^2}} dv,
    \end{split}
\end{equation}
    \begin{equation}\label{eq: explicitIonApp}
    \begin{split}
         &\textbf{L}_{\textbf{ion}} =  \int^\infty_{-\infty} \mathcal{N}\left(\mu, \sigma^2\right)  dy \int^{2\pi}_0 d \varphi\int^\frac{\pi}{2}_0 \sin{2 \theta} 
       \\ & \times \int^\infty_0 (1-\lambda_i)
       \frac{1.4132 \pi^3\beta_i \beta^2_{ge} I_0^3 v^2 w^2}{\sqrt{\pi} u^3 v_\perp^2}
    \exp{-\frac{6\rho^2}{w^2} - \frac{D}{\tau v_\perp} -\frac{ v^2}{u^2}} V \left(\delta f + \Delta f_{AC} - \frac{v^2}{2c^2} f_0;1.1132 \eta ;  0.4535 \eta\right) dv.
    \end{split}
\end{equation}
\end{widetext}
The functions $\textbf{L}_{\textbf{2S}}$ and  $\textbf{L}_{\textbf{ion}}$ are the final functions we derive in this appendix. Figure \ref{fig:Lion} in the main text presents a plot of $\textbf{L}_{\textbf{ion}}$ alongside the full Monte Carlo simulation of the experiment. As can be seen, the two are in excellent agreement at low power (46\,W), where perturbation theory is highly reliable, but $\textbf{L}_{\textbf{ion}}$ slightly overestimates the ionisation probability close to resonance at high power (460\,W) — which is to be expected from this perturbative treatment.

The lineshape model accounts for transit-time broadening, the second-order Doppler effect, the finite lifetime of the 1S state, photoionization, Ps emission distributions, and the AC Stark shift. These are the dominant physical mechanisms that affect the shape of the spectrum. It does not, however, include the finite lifetime of the 2S state, and it treats the beam waist as constant rather than allowing it to vary hyperbolically, as would be expected for a Gaussian beam. In our experimental setup, these omissions are expected to have only a very minor effect on the lineshape, so neglecting them is well justified.

With some minor adjustments, these lineshape functions could be straightforwardly adapted for $\text{1}^\text{3}\text{S}_\text{1} \to \text{2}^\text{3}\text{S}_\text{1}$ spectroscopy of other unstable simple atomic systems, such as muonium.

\section{\label{sec:appendix_frequency_metrology} Frequency metrology}
The laser frequency used to excite the $\text{1}^\text{3}\text{S}_\text{1} \to \text{2}^\text{3}\text{S}_\text{1}$ transition in positronium was measured and controlled using a HighFinesse WS-7 wavemeter. The PID output of the device acts on the piezoelectric element voltage of the laser ECDL's grating to stabilise frequency. The typical daily frequency drifts of the wavemeter is on the order of 10 MHz, attributed to thermal and pressure variations in the laboratory environment. To account for this drift, the device is regularly calibrated against the rubidium $5^2S_{1/2}(F = 2) \rightarrow 5^2P_{1/2}(F = 3)$ transition at 780.2 nm (384.227 848 THz). The transition has a linewidth of 6 MHz, provides an absolute reference with sub-MHz accuracy. The rubidium standard, located in a neighbouring laboratory, is delivered via a stabilised optical fibre to the experiment. Calibration of the wavemeter against this standard allows the laser frequency to be known with a relative accuracy of 1 MHz. During the spectroscopy data taking, the wavemeter is periodically calibrated against the rubidium reference every 8 hours. The wavemeter is expected to drift between calibrations, and discrepancy between the actual laser frequency and the nominal value provided by the wavemeter is determined by linearly interpolating between successive calibration measurements.
\begin{figure}[h!]
\includegraphics[width=\columnwidth]{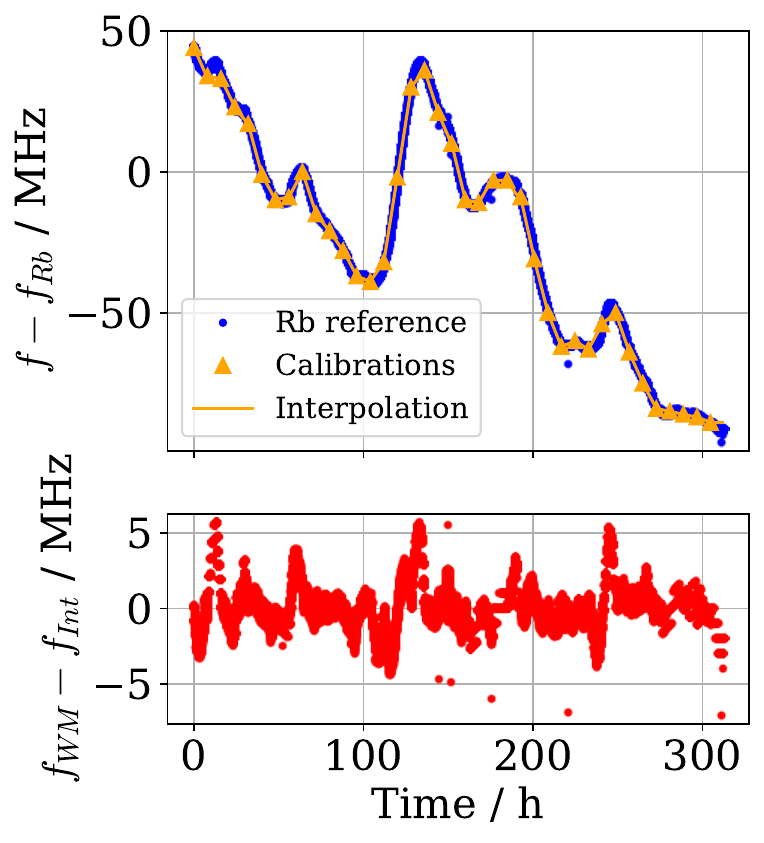} 
   \caption[Continuous measurement of the rubidium reference by the wavemeter over a timescale of thirteen days before measurement.] {Continuous measurement of the rubidium reference by the wavemeter over a timescale of thirteen days before measurement. In the top plot the zero level corresponds to the expected reference frequency. The blue datapoints indicate the frequency measured by the wavemeter. Measurements jumping far from the trend are attributed to the sparse unlocking of the reference laser. The orange datapoints represent the expected wavemeter calibrations when running spectroscopy experiments, with linear interpolations between them. The red points in lower plot displays the residuals between the wavemeter measurements and linear interpolation, indicative of the frequency errors expected when wavemeter calibration method described here is employed.
\label{fig:wavemeter drift}}
\end{figure}

To model the frequency stability resulting from calibration and linear interpolation, the rubidium reference transition was continuously measured over a period of 13 days, as illustrated in Fig. \ref{fig:wavemeter drift}. The rubidium standard provides a constant frequency reference, so the frequency drift shown in Fig. \ref{fig:wavemeter drift} originates from drift in the wavemeter measurement. To simulate the effects of regular calibration and interpolation, every 8 hours a wavemeter reading in the data set displayed in Fig. \ref{fig:wavemeter drift} is selected as a calibration point, which are displayed in orange, and these points are linearly interpolated between. The residuals from the linear interpolations, displayed below in red, represents the frequency error from this method.

\begin{figure}[h!]
\includegraphics[width=\columnwidth]{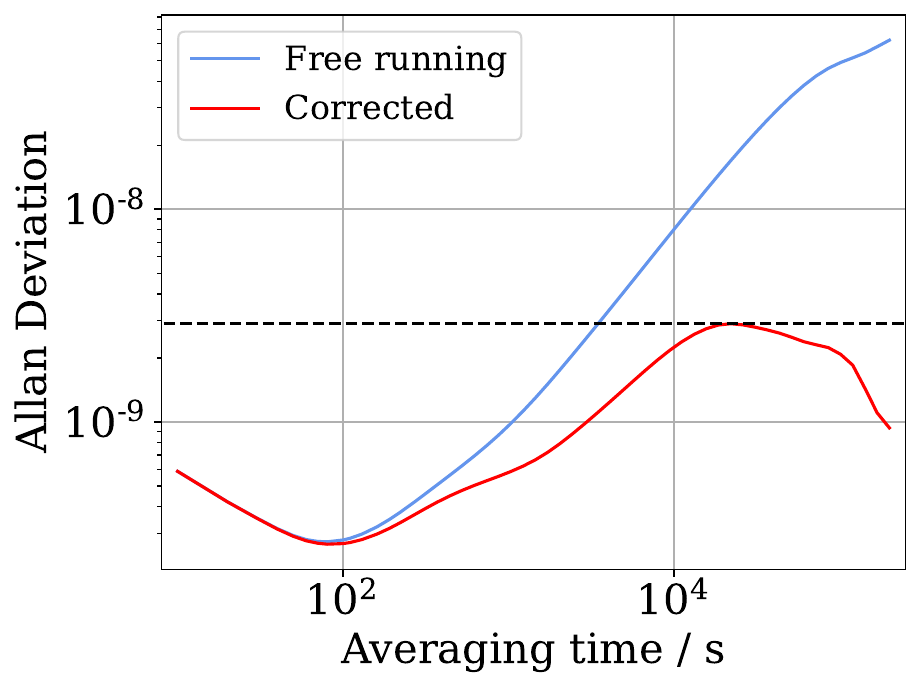} 
   \caption{Wavemeter drift uncertainty calculated from Allan deviation. A linear interpolation approximates well the behaviour over the plotted timescale.
\label{fig:Allan_dev}}
\end{figure}

To estimate the frequency uncertainty arising from this approach, we generate Allan deviation plots, shown in Fig. \ref{fig:Allan_dev}, produced using the red and blue frequency data in Fig. \ref{fig:wavemeter drift}. After about $10^2$s the Allan deviation increases, which indicates a random walking or drifting frequency. However, the Allan curve of the red dataset, which represents the frequency stability resulting from regular calibration and interpolation, shows a more gradual increase in Allan deviation compared to the curve for the uncalibrated blue dataset.
The red Allan curve peaks at just below 8 hours, which reflects how the frequency uncertainty is at is highest just before calibration. As a conservative estimate of the uncertainty that results from the calibration and interpolation method, we take the value of the Allan deviation at the maxima, which is indicated by the horizontal dashed line and has a value of $2.9 \times 10^{-9}$, which corresponds to 3.58\,MHz at 243\,nm.

An additional systematic shift of the wavemeter reading was characterized when calibrated at 780.2nm and used to measure the laser at 972nm. This wavelength offset leads to a systematic underestimation of the frequency by -14.2(0.7) MHz, which was corrected in the final frequency determination. Given that the 1S–2S transition involves two photons at 486nm, this systematic offset need to be multiply by 4 and contributes a total uncertainty of approximately -56.8(2.8) MHz to the measured transition frequency.

\section{\label{sec:appendix_DCStarkShift} DC Stark shift}

This appendix outlines the calculation of the static dipole polarisability and DC Stark shift for the $1S$ and $2S$ states of positronium in a uniform electric field. The treatment follows standard non-relativistic perturbation theory in atomic units and provides quantitative estimates of the level shifts relevant for precision spectroscopy of the $1S$–$2S$ transition and the uncertainty in this systematic shift arising from the knowledge of the field in the excitation region.

Calculations are performed in atomic units (a.u.), defined by $\hbar = m_e = e = a_0 = 1$. In these units, the (static) dipole polarisability $\alpha$ has units of volume ($a_0^3$). For an electric field $\mathcal{E}$ in SI units,
\begin{equation}
    E_{\mathrm{a.u.}} = \frac{\mathcal{E}}{E_0}, \qquad
    E_0 = \frac{E_h}{e a_0} \approx 5.1422067\times 10^{11}\ \mathrm{V/m},
\end{equation}
where $E_h$ is the Hartree energy. All energy shifts are ultimately expressed as frequency shifts through $\Delta\nu = \Delta E / h$. For positronium, the Bohr radius is $a_{\mathrm{Ps}} = 2 a_0$ due to the reduced mass $\mu = m_e/2$. Energies in atomic units are converted to hertz using $E_h/h$.

A static, uniform electric field $\mathbf{E} = \mathcal{E}\, \hat{\mathbf{z}}$ introduces the perturbation
\begin{equation}
V = -\mathcal{E}\, z.
\end{equation}
For a non-degenerate initial state $\ket{i}$, the second-order energy correction is
\begin{equation}
\Delta E^{(2)} = -\frac{1}{2}\,\alpha\,\mathcal{E}^2, \qquad
\alpha = 2 \sum_{n} \frac{|\matrixel{i}{z}{n}|^2}{E_n - E_i},
\label{eq:alpha-def}
\end{equation}
where the sum extends over all intermediate states with opposite parity ($P$-states for $S$ levels).

\subsection{Polarisability of the 1S State}

For positronium, the wavefunctions are hydrogenic with Bohr length $a_{\mathrm{Ps}} = 2a_0$. The dipole polarisability of an $nS$ state scales as $a_{\mathrm{Ps}}^3$. The analytical result for the $1S$ state is
\begin{equation}
   \alpha_{1S} = \frac{9}{2}\, a_{\mathrm{Ps}}^3 = 36\, a_0^3.
\end{equation}

The corresponding Stark shift is
\begin{equation}
\Delta E_{1S}^{\mathrm{Stark}} = -\tfrac{1}{2}\, \alpha_{1S}\, \mathcal{E}^2,
\qquad
\Delta \nu_{1S}^{\mathrm{Stark}} = -\tfrac{1}{2}\,\alpha_{1S}\,E_{\mathrm{au}}^2\,\frac{E_h}{h}.
\end{equation}
For the electric field $\mathcal{E} = \SI{340}{V/m}$ used in our experiment,
\begin{equation}
   \Delta \nu_{1S}^{\mathrm{Stark}} = \SI{-0.05}{Hz}.
\end{equation}
This effect is negligibly small compared with the $2S$ Stark shift discussed below.

\subsection{2S Stark Shift from $2^3P_J$ Mixing}

The $2^3S_1$ level is coupled by the electric dipole interaction $-e \mathcal{E} z$ to the nearby $2^3P_J$ manifold ($J=0,1,2$). The dominant Stark shift arises from this near-resonant mixing. Second-order perturbation theory gives
\begin{equation}
\Delta E_{2S}^{\mathrm{Stark}}
= -\sum_{J}
\frac{\left|\matrixel{2\,^3P_J m_J}{e\mathcal{E} z}{2\,^3S_1 m_J}\right|^2}
     {E_{2\,^3P_J}-E_{2\,^3S_1}}.
\end{equation}

The line strengths scale with $(2J+1)$, giving normalized weights
\begin{equation}
w_0:w_1:w_2 = 1:3:5, \qquad
w_0 = \tfrac{1}{9},\; w_1 = \tfrac{3}{9},\; w_2 = \tfrac{5}{9}.
\end{equation}
For $n=2$ hydrogenic states, the reduced radial matrix element is
\begin{equation}
    \big| \matrixel{2S}{e z}{2P} \big| = 3 e a,
\end{equation}
which for positronium becomes $|d| = 3 e a_{\mathrm{Ps}}$.

Defining the fine-structure splittings
\begin{equation}
\Delta \nu_J \equiv \frac{E_{2^3S_1} - E_{2^3P_J}}{h} > 0,
\end{equation}
the total shift of the $2S$ level is
\begin{equation}
\Delta \nu_{2S}^{\mathrm{Stark}}
= -\left(\frac{3 e a_{\mathrm{Ps}} \mathcal{E}}{h}\right)^{\!2}
\sum_{J=0}^2 \frac{w_J}{\Delta \nu_J}.
\label{eq:2Sstark}
\end{equation}
Since $\Delta\nu_J > 0$, the shift is a red shift.

\subsection{Numerical Estimate}

Using the experimental $2S$–$2^3P_J$ splittings
\[
\Delta \nu_{J=2} = \SI{8.6}{GHz}, \quad
\Delta \nu_{J=1} = \SI{13.0}{GHz}, \quad
\Delta \nu_{J=0} = \SI{18.5}{GHz},
\]
and a field $\mathcal{E} = \SI{340}{V/m}$, one finds
\begin{align}
E_{\mathrm{au}} &= \frac{340}{E_0} \approx 6.61\times 10^{-10}, \\[2mm]
\Delta \nu_{1S}^{\mathrm{Stark}} &= -5.2\times 10^{-2}\ \mathrm{Hz}, \\[2mm]
\Delta \nu_{2S}^{\mathrm{Stark}} &= -6.6\times 10^{4}\ \mathrm{Hz}.
\end{align}
Thus the total shift of the $1S$–$2S$ transition is
\begin{equation}
\Delta \nu_{2S-1S}^{\mathrm{Stark}} \approx -\SI{66}{kHz}.
\end{equation}
Assuming a residual-field uncertainty of $\SI{10}{V/m}$, the resulting uncertainty in the Stark correction is about $\SI{4}{kHz}$.

\bibliographystyle{apsrev4-2}
\bibliography{literature}


\end{document}